\documentclass[a4paper]{JHEP3} 

\usepackage{bbm,amsmath,graphicx,amssymb}
\usepackage{bm}
\usepackage{cite}
\usepackage{verbatim}
\usepackage{slashed}
\usepackage{multirow}

\newcommand{\bea}{\begin{eqnarray}}
\newcommand{\eea}{\end{eqnarray}}
\newcommand{\bean}{\begin{eqnarray*}}
\newcommand{\eean}{\end{eqnarray*}}

\def\Label#1{\label{#1}%
  \smash{\hbox to0pt{\raise1ex\hbox{\tiny[#1]}\hss}}}

\def\stamp{--- {\bf \today} --- {\bf \jobname.tex}}

\def\fs_#1{\mathfrak{s}(#1)}

\def\BE{\begin{equation}}
\def\EE{\end{equation}}
\def\spa#1.#2{\left\langle#1\,#2\right\rangle}
\def\spb#1.#2{\left[#1\,#2\right]}
\def\lor#1.#2{\left(#1\,#2\right)}

\newcommand\fverb{\setbox\fverbbox=\hbox\bgroup\verb}
\newcommand\fverbdo{\egroup\medskip\noindent%
            \fbox{\unhbox\fverbbox}\ }
\newcommand\fverbit{\egroup\item[\fbox{\unhbox\fverbbox}]}
\newbox\fverbbox

\title{The Complete KLT-Map Between Gravity and Gauge Theories}

\author{Poul H. Damgaard, Rijun Huang, Thomas S{\o}ndergaard, Yang Zhang\\
Niels Bohr International Academy and Discovery Center,\\ The Niels Bohr
Institute,
Blegdamsvej 17,\\ DK-2100 Copenhagen \O, Denmark\\ {\tt emails:}
\{phdamg, huang, tsonderg, zhang\}@nbi.dk}

\received{\today}       
\accepted{\today}       

\abstract{We present the complete map of any pair of super Yang-Mills
theories to supergravity theories as dictated by the KLT relations in
four dimensions. Symmetries and the full set of associated vanishing
identities are derived. A graphical method is introduced which simplifies
counting of states, and helps in identifying the relevant set of symmetries.}


\begin{document}
\section{Introduction}

While general relativity has a beautiful and compact formulation
in terms of the Einstein-Hilbert action, it is striking that all
simplicity is lost once attempts to view that same action from
the perturbative (graviton) point of view. The simple and almost
trivial problem is that there is no way to
treat both the metric $g_{\mu\nu}$ and its inverse as a background
with a small single-term perturbative correction. As a consequence,
the Einsten-Hilbert action explodes into an infinite
series of terms when expanded about a fixed background: There is
an infinite series of vertices of increasing order in the number
of graviton fields. Beyond four-point scattering amplitudes this
makes it essentially impossible to compute tree-level graviton
scattering amplitudes on the basis of conventional Feynman-diagram
techniques.

A most surprising and powerful way to circumvent this
obstacle to perturbative calculations in gravity (or theories coupled
to gravity) is provided by the Kawai-Lewellen-Tye (KLT) relation
\cite{KLT}. Based on the factorization of closed-string amplitudes
into a product of two open-string amplitudes, it yields, in the
field theory limit, a remarkable connection between gravity and
Yang-Mills theory. This is totally obscure at the level of
the actions of gravity and Yang-Mills theory.
Symbolically, it tells us that gravity, as far as tree-level $n$-point
amplitudes are concerned, is a very particular kind
of 'square' of Yang-Mills theory,
$$
Gravity \sim (Gauge ~theory)\times(Gauge ~theory)\,.
$$
This is just the field theory limit of the KLT-relation, and because
of its existence in the full superstring case, it really, more
generally, relates supergravity theories to super Yang-Mills theories.
The precise map in field theory has recently been derived by means of
on-shell recursion for all $n$-point functions \cite{BjerrumBohr1},
\bea
M_n = \sum_{\gamma,\beta\in S_{n-3}}
\widetilde{A}_{n}(n-1,n,\gamma,1) \mathcal{S}[\gamma|\beta]_{p_1}
A_{n}(1,\beta,n-1,n)
\eea
Here, $M_n$ indicates an $n$-point gravity amplitude, and on the right
hand side we have Yang-Mills $n$-point amplitudes (the possibility
of combining two different kinds of Yang-Mills amplitudes $A_n$ and
$\widetilde{A}_n$ will be discussed in details below). The object
$\mathcal{S}$ that
glues the two Yang-Mills amplitudes together, the S-kernel
\cite{BjerrumBohr1}, will be defined below. It serves in a very
precise manner to remove the double poles of the amplitude product
and, simultaneously, to ensure full permutation symmetry of the
gravity amplitude. Note that in the sum above only permutation symmetry
of $n-3$ of the $n$ legs is manifest. With a very simple
modification, the S-kernel of field theory generalizes to the
full string theory case \cite{BjerrumBohr2}.

In string theory, the factorization of closed-string amplitudes into
two open-string amplitudes is not unique \cite{KLT}. Indeed, the
S-kernel that ties the two Yang-Mills amplitudes together is also
not unique: there is a whole family of kernels that all do the factorization
correctly. Conversely, this must imply amplitude relations on
the Yang-Mills side. This intuition turns out to be correct:
the Bern-Carrasco-Johansson (BCJ)
relations that were first conjectured in field theory \cite{BCJ}
and later shown to follow from monodromy in string theory
\cite{BjerrumBohr3,Stieberger}, are intimately tied to the
KLT-relations\footnote{See also \cite{Ma:2011um}.}. The BCJ-relations have also, like the KLT-relations, been derived
in field theory by means of on-shell recursion \cite{BoRijun,Chen:2011jxa}, and the S-kernel of the
KLT-map can be seen as a generator of BCJ-relations
\cite{BjerrumBohr1,BjerrumBohr2}.
A variant of BCJ-relations that builds on numerators of
given amplitude representations \cite{BCJ} constructs amplitudes
for gravity by squaring numerators. This has been proven
to give a remarkable alternative representation of gravity amplitudes
by means of products of two Yang-Mills amplitudes \cite{Bern1}.
Perhaps the most exciting aspect is that
it points directly towards applications of KLT squaring relations
also at loop level \cite{Bern2}. There are also applications
of KLT relations directly at one-loop order \cite{Naculich:2011my}.

A common theme in these recent developments is the appearance of
surprising traces of Yang-Mills theory in perturbative gravity. From
scattering amplitudes it is as if the color group is replaced by a
kinematical group. This is an idea that originates in the way Jacobi
relations of the color group in Yang-Mills theory have (generalized)
mirrors  in the kinematical factors that appears in numerators
\cite{BCJ,TyeZhang,BjerrumBohr4,Bern3,Stieberger1}. These notions
have become beautifully synthesized in recent work of Monteiro and
O'Connell \cite{Donal}, where, conversely, the diffeomorphism
invariance of gravity re-surfaces in the kinematics of Yang-Mills
theory. The connection between BCJ-relations and the weak-weak
duality between gauge theory and gravity is more and more leading us
towards an {\em algebra of amplitudes} \cite{BjerrumBohr5}. This
serves as additional motivation to establish the precise
relationship between two (super) Yang-Mills theories and the
associated (super) gravity theory as dictated by the KLT-map.
Several aspects of this gauge theory to gravity map can already be
found, scattered through the scientific literature, in terms of
specific examples. Here, we provide the comprehensive catalog of the
maps between four-dimensional gauge theories with supersymmetries
${\mathcal N}= 4, 3, 2, 1$ and 0 to the corresponding supergravity
theories. The ${\mathcal N}=0$ theory is nothing but pure Yang-Mills
theory. Yet, as is well known, it maps not to pure Einstein gravity,
but to gravity coupled to two real scalars: in string theory language
an axion-dilaton pair.

Indeed, the recent field-theory proof of
KLT-relations \cite{BjerrumBohr1} hinged in an essential manner on a
corresponding set of ``vanishing relations" \cite{BjerrumBohr6},
quadratic identities among Yang-Mills amplitudes that were proven
independently. These vanishing identities correspond to gravity
amplitudes with an odd number of (complex) scalars, which vanish. As
observed in \cite{TyeZhang1}, such identities indeed follow from
string theory. Alternatively, if seen as embedded in maximally supersymmetric theories,
they correspond to KLT-relations that violate $R$-symmetry on the gravity side \cite{FengHe,Elvang}. It is
interesting that even if one considers only the conventional
KLT-map of vector bosons to gravitons with like helicities, the scalars
never appear. Yet, as we have mentioned, they are essential for
the proof \cite{BjerrumBohr1} of these KLT-relations {\em even in the
complete absence of scalars as external states.} Much can be
learned already from this  simplest example, since it shows that
by gluing two pure Yang-Mills theories together through the S-kernel
of the KLT-map, we do not get just pure gravity, and that the
additional states on the gravity side are crucial for understanding
the gauge theory to gravity map. Let us explain this in slightly
greater detail.

We start by considering gravity as the low-energy limit of
string theory. It consists of the following states and their
corresponding polarization tensors:
\begin{itemize}
\item Graviton. $e_{\mu\nu}^{++}(k)=\epsilon_\mu^+(k)
  \epsilon_\nu^+(k)$\,, \quad
  $e_{\mu\nu}^{--}(k)=\epsilon_\mu^-(k)  \epsilon_\nu^-(k)$\,.
\item Axion.  $e_{\mu\nu}(k)=\epsilon_\mu^+(k)
  \epsilon_\nu^-(k)-\epsilon_\mu^-(k)  \epsilon_\nu^+(k)$\,.
\item Dilaton.  $e_{\mu\nu}(k)=\epsilon_\mu^+(k)
  \epsilon_\nu^-(k)+\epsilon_\mu^-(k)  \epsilon_\nu^+(k)$\,.
\end{itemize}
Gravity amplitudes with axions and dilations have conserved
quantum numbers, and these constraints lead to vanishing
identities when combined with KLT-relations. To see the origin of these
identities, let us present the action of this theory \cite{Schwarz:1992tn}. The four-dimensional
coupled axion-dilaton gravity action reads as follows in the Einstein frame:
\begin{equation}
  S=\frac{1}{2 \kappa^2}\int d^4 x \sqrt{-G} (R- 2\partial_\mu \phi \partial^\mu \phi
-\frac{1}{12} e^{-4 \phi} H_{\mu \nu \rho} H^{\mu \nu \rho})\,.
\end{equation}
The Poincare dual of $H_{\mu \nu \rho}$ is the axion,
\begin{equation}
\partial_\mu
b=\frac{1}{6} e^{-4 \phi} \epsilon_{\mu \nu \rho \sigma} H^{\mu \nu
  \rho} ~.
\end{equation}
We can now combine the axion and the dilaton into the complex
combination $z = b + i e^{-2 \phi}$, leading to
\begin{equation}
  S=\frac{1}{2 \kappa^2}\int d^4 x \sqrt{-G} \left( R- \frac{1}{2} \frac{|\partial_\mu z|^2}{(\text{Im} z)^2}\right)\,.
\end{equation}
Here, $z$ takes value in the upper complex plane, as can be seen from its definition.
This upper part of the complex plan is the moduli space
of this theory, which  has an $SL(2, \mathbb{R})$ global symmetry:
\begin{equation}
 g_{\mu\nu} \mapsto g_{\mu\nu}\,, \quad  z \mapsto \frac{a z +b }{c z +d }\,,
\quad a d - b c=1\,.
\end{equation}
We may choose the vacuum expectation value as $\langle z \rangle =i$.
Then the remaining manifest symmetry is $U(1)$,
\begin{equation}
 \begin{pmatrix}
  a & b  \\
  c & d  \\
 \end{pmatrix}
=\begin{pmatrix}
  \cos \theta & \sin \theta  \\
  -\sin \theta & \cos\theta  \\
 \end{pmatrix}\,.
\label{SL(2,R)_unbroken}
\end{equation}
We can also perform a redefinition of the scalar field in order
to make the symmetry linearly realized.
This is achieved by
\begin{equation}
z=\frac{ 2 \kappa w}{1+i \kappa w}+i ~,
\end{equation}
which is a M\"obius transformation which transforms the upper half plane to the
Poincare disc,
$  |w| < 1/\kappa$. The action then becomes
\begin{eqnarray}
  S=\frac{1}{2 \kappa^2}\int d^4 x \sqrt{-G} R - \int d^4 x \sqrt{-G}
\frac{\partial_\mu w \partial^\mu \bar w}{(1-  \kappa^2 |w|^2 )^2}\,,
\label{action-disc}
\end{eqnarray}
and the $U(1)$ symmetry acts on $w$ as $w \mapsto e^{2 i\theta}
w$. This charges the states $\epsilon_\mu^+(k)
  \epsilon_\nu^-(k)$ and $\epsilon_\mu^-(k)
  \epsilon_\nu^+(k)$. The origin of the vanishing identities as arrived
from KLT-relations is
now clear. The action of the remaining two generators of $SL(2,\mathbb{R})$ changes
the vacuum expectation value of $z$. The axion and the dilaton can thus
be regarded as the two Goldstone bosons associated with the global
symmetry breaking $SL(2,\mathbb{R}) \sim SU(1,1) \to U(1)$. We will see analogs
of these two Goldstone bosons in all the cases to be discussed below.
In the maximally supersymmetric case, the scalars  live in the well-known
coset of $E_{7(7)}/SU(8)$.

The pure gauge theory case discussed above gives a very direct proof of
the vanishing identities.
However, as stressed in ref. \cite{Elvang}, it can also be profitable to understand
these identities from violation of $R$-symmetry in the maximally supersymmetric
version of the KLT-map. This may appear puzzling, since supersymmetry
plays no direct role
in the identities themselves. However, in taking such a top-down approach, one sees
every single four-dimensional  KLT-map between gauge theory and gravity as a sub-map
of the maximal map between two ${\mathcal N}=4$ super Yang-Mills theories to
${\mathcal N}_G=8$ supergravity. Crucial in this connection is a most powerful extension
of the proof of the KLT-formula in terms of superfields by Feng and He \cite{FengHe}.
We shall here use this extended KLT-formula to project out more and more
fields from the maximally supersymmetric case, in this way generating a complete catalog
of KLT-maps between theories with
less and less supersymmetry in four dimensions. The ${\mathcal N}=4$ superfield of
Nair \cite{Nair} and the associated definition of a {\em superamplitude} out of given
helicity amplitudes, is essential in this context.
For each entry in the KLT-table we identify global
symmetries including $R$-symmetries, if applicable,
 and these symmetries determine
the set of vanishing identities.

Our paper is organized as follows. In section 2 we briefly review
the KLT-relation and its generalization to maximal supersymmetry in its superfield
formulation. In
section 3 we discuss the KLT-relations in superfields for $\mathcal{N}<4$
super Yang-Mills theories, and describe the full set of supergravity theories that can be
obtained from KLT products. As an aid in the construction, we introduce a graphical tool
(``diamond diagrams") for KLT-relations of non-maximal theories. In section 4 we study
the invariant symmetry groups that emerge naturally from KLT
products for the various supergravity  theories. Conclusions and an outlook are given in
the final section.

\section{A brief overview of KLT-relations}

The full, stringy, KLT-relation \cite{KLT} has recently been shown to
take the following explicit form \cite{BjerrumBohr2}\footnote{The relation that is derived
from the low-energy limit of string theory comes with an overall irrelevant sign 
factor \cite{BjerrumBohr2} which we ignore in this paper.}:
\bea
M_n^{closed}= \sum_{\gamma,\beta}\widetilde{A}_n^{open}(n-1,n,\gamma,1)
{\mathcal S}_{\alpha'}[\gamma|\beta]_{p_1}A_n^{open}(1,\beta,n-1,n)
~,~~~\label{string-KLT}\eea
where we sum over two sets of $(n-3)$ permutations $\beta$ and
$\gamma$. The momentum kernel ${\mathcal S}_{\alpha'}$ glues two open string-theory
amplitudes $A_n^{open}, \widetilde{A}_n^{open}$ together to form the closed string-amplitude
$M_n^{closed}$. The explicit form of ${\mathcal S}_{\alpha'}$, given in ref.
\cite{BjerrumBohr2}, resembles very much its field theory analog
$\mathcal{S}$, which will be defined below. The two open string
amplitudes may be bosonic or supersymmetric depending on whether
the closed string amplitude is bosonic or supersymmetric.
Alternatively, one of the open string amplitudes may be bosonic and
the other supersymmetric if one wishes to construct a heterotic
string amplitude\footnote{Some explicit examples can be found in refs.
\cite{Bern6}.}.  This implies that in the field theory
limit, we can either have KLT-relations between gravity and
Yang-Mills theory, or we can have relations between supergravity and
super Yang-Mills theories. The $\alpha'$-dependent momentum kernel
${\mathcal S}_{\alpha'}$ is totally independent of the types of closed and
open strings considered, and thus the relation (\ref{string-KLT}) is
completely general and universal. In the field theory limit of
$\alpha' \to 0$, the momentum kernel ${\mathcal S}_{\alpha'}$ reduces
directly to the field theory S-kernel.
Therefore also in field theory the S-kernel
is universal, and independent of whether the involved Yang-Mills and
gravity amplitudes arise in supersymmetric theories or not.

It is clear, then, that in the field theory limit, the KLT-relations express
a given gravity amplitude as a particular sum of gauge field amplitudes
squared.
An explicit expression of this relation with manifest $(n-3)!$
permutation symmetry has been proven to be \cite{BjerrumBohr1}
\bea
M_n = \sum_{\gamma,\beta\in S_{n-3}}
\widetilde{A}_{n}(n-1,n,\gamma,1) \mathcal{S}[\gamma|\beta]_{p_1}
A_{n}(1,\beta,n-1,n)~,~~~\label{non-susy-KLT}\eea
where $\gamma$ and $\beta$ are permutations over the legs
$2,\ldots,n-2$ and the S-kernel is defined by
\bea  \mathcal{S}[i_1,\ldots,i_m|j_1,\ldots,j_m]_{p_1} \equiv
\prod_{t=1}^m(s_{i_t1} + \sum_{q>t}^m \theta(i_t,i_q)
s_{i_ti_q})~,~~~\label{s-kernel}\eea
with
\bean \theta(i_a,i_b) \equiv \left\{
\begin{array}{ll}
  1 & \text{   if } i_a \text{ appears \textbf{after} } i_b \text{ in the sequence } \{j_1,\ldots,j_m\} \\
 0 & \text{   if } i_a \text{ appears \textbf{before} } i_b \text{ in the sequence } \{j_1,\ldots,j_m\}
\end{array}
\right.~.~~~ \eean
The shown expression is not unique: there is a whole class of different but
equivalent KLT-relations written in terms of S-kernels \cite{BjerrumBohr1,BjerrumBohr2}. One of these,
the one containing fewest terms,
coincides with an explicit representation conjectured already in Appendix A of ref. \cite{Bern4}.

As discussed in the introduction, the KLT-formula \eqref{non-susy-KLT} is quite surprising
since the gravity amplitude on the left hand side receives contributions from Feynman diagrams
with, depending on the number of external legs $n$, graviton vertices that
continue up to infinite order while the Yang-Mills amplitudes on the right hand side
of course receive contributions of only three- and four-point vertices.
Another surprising aspect of the KLT-relation is that Yang-Mills amplitudes
are colored objects, but gravity should know nothing about color. Magically, it is only
the color-stripped Yang-Mills amplitudes that enter. But these
are not symmetric under permutations of the external legs, while of course the
gravity amplitude is totally symmetric under such
permutations. Though not obvious, the right hand
side of \eqref{non-susy-KLT} is indeed symmetric under permutations of all $n$ legs,
rather than just the manifest symmetry under $(n-3)!$ of such permutations.
Finally, there is magic with respect to locality. A simple Feynman-diagram analysis
quickly shows that the product of two gauge theory amplitudes has double poles
that cannot be allowed in a gravity amplitude. The S-kernel cleverly manages to precisely
cancel these unwanted double poles, rendering the correct behavior required for
the gravity amplitude on the left hand side. A crucial series of very finely tuned
cancellations clearly occur in the KLT-relation. It is from this point of view even more
striking that the S-kernel is not unique, and that a whole series of such kernels can
do the job. This is precisely the origin of the BCJ-relations, which can be viewed as
a consequence of the equivalence between these different parametrization of the
KLT-relations.\footnote{A first example of this, at the six-point level, was already implicit
in the original KLT-paper \cite{KLT}.} Indeed, as also mentioned in the introduction, the S-kernel
can be regarded as the generator of BCJ-relations \cite{BjerrumBohr1,BjerrumBohr2}:
\bea
\sum_{\sigma\in S_{n-2}} {\mathcal S}
[\sigma_{2,n-1}|j_2,\ldots,j_{n-1}]_{p_1}A_n(n,\sigma_{2,n-1},1)
~=~ 0 ~.
\eea
For a more detailed review of the above see \cite{Sondergaard}.

In order to compactly phrase the KLT-relations for supersymmetric theories
with a possible large set of different external states, it is convenient
to introduce the superfield description of those relations \cite{FengHe,Elvang}.
For maximally supersymmetric theories, we arrange all
on-shell component fields {\em
listed according to their helicities} into a
single superfield \cite{Nair}. We recover all component fields by
an expansion of the superfield in the associated Grassmann variables $\eta$.
The particle states of $D=4$,  $\mathcal{N}=4$ super Yang-Mills theory are
two gluons $g_+, g_-^{1234}$, four pairs of gluinos $f_+^a,
f_-^{abc}$ and six scalars $s^{ab}$ satisfying the reality
condition $s^{ab} = \epsilon^{abcd}s_{cd}/2$, where $s^{ab} \equiv s_{ab}^{\dagger}$.
They transform under
$SU(4)_R$ symmetry as anti-symmetric products in the fundamental
representation and therefore carry the anti-symmetric fundamental
$SU(4)_R$ indices $a,b,\cdots = 1,2,3,4$. We pack them together
into an $\mathcal{N}=4$ superfield
\bea
\Phi^{\mathcal{N}=4} = g_+ + \eta_a f_+^a +
\frac{1}{2!}\eta_a\eta_b s^{ab} + \frac{1}{3!}\eta_a\eta_b\eta_c
f_-^{abc} +\eta_1\eta_2\eta_3\eta_4
g_-^{1234}~,~~~\label{Phi-N4}
\eea
where the $\eta_a$'s are Grassmann variables labeled by $SU(4)_R$
symmetry indices. In the on-shell formalism the supercharges are
given by
\bea
\widetilde{Q}_a = \sum_{i=1}^{n}| i
\rangle \eta_a~,~~~Q^a = \sum_{i=1}^n|
i\rbrack \frac{\partial}{\partial \eta_a}~,~~~
\eea
which relates all the 16 states in one supermultiplet.

It is useful to think of the $\Phi$'s as super-states and
introduce a superamplitude
$$\mathcal{A}_n^{\mathcal{N}=4}(\Phi_1,\Phi_2,\ldots,\Phi_n)\,,$$
which represents a sum of amplitudes of all different helicity assignments
and choices of external states. The expansion coefficients, which
uniquely identify a given component helicity amplitude, are
precisely the $\eta_a$'s, one set for each external line.
Because the amplitudes must be invariant under $SU(4)_R$
symmetry, this puts constraints on the combinations of indices that
can occur for non-vanishing amplitudes. Hence many of the amplitudes in
the direct expansion will vanish since they are $SU(4)_R$ symmetry
violating. Schematically, what is left is thus an expansion of the form
\begin{align}
\mathcal{A}_n^{\mathcal{N}=4} = \sum A_n^{MHV} (\eta)^8 + \sum
A_n^{NMHV} (\eta)^{12} + \ldots + \sum A_n^{\overline{MHV}}
(\eta)^{4n-8}\,, \label{A_exp}
\end{align}
where each $SU(4)_R$ symmetry index ($a=1,2,3,4$) appears the same
number of times in each monomial of the $\eta_a$'s. Here
$A_n^{N^kMHV}$ denotes the actual component helicity amplitudes.
They can be extracted from the superamplitudes by acting with the
corresponding differential operators (or integrals) that single out the
desired components.

The $\mathcal{N}_G=8$ supergravity theory has an on-shell
formalism that is completely analogous to $\mathcal{N}=4$ super Yang-Mills theory.
The superfield of $\mathcal{N}_G=8$
supergravity contains one graviton $h_{\pm}$, 8 gravitinos $\psi_{\pm}$, 28
graviphotons $v_{\pm}$, 56 graviphotinos $\chi_{\pm}$ and 70 real
scalars $\phi$. It can be represented as
\bean
\Phi^{\mathcal{N}_G=8} ={}& h_+ + \eta_A \psi_{+}^A +
\frac{1}{2!}\eta_A\eta_B v_+^{AB} + \frac{1}{3!}\eta_A\eta_B\eta_C
\chi_+^{ABC}
+ \frac{1}{4!}\eta_A\eta_B\eta_C\eta_D \phi^{ABCD} \nonumber \\
& + \frac{1}{5!} \eta_A\eta_B\eta_C\eta_D\eta_E \chi_-^{ABCDE} +
\frac{1}{6!}\eta_A\eta_B\eta_C\eta_D\eta_E\eta_F v_-^{ABCDEF} \nonumber \\
& + \frac{1}{7!} \eta_A\eta_B\eta_C\eta_D\eta_E\eta_F\eta_G
\psi_-^{ABCDEFG} + \eta_1\eta_2\eta_3\eta_4\eta_5\eta_6\eta_7\eta_8
h_-^{12345678}~,~~~\label{Phi-N8}
\eean
and we can likewise introduce superamplitudes for $\mathcal{N}_G=8$
supergravity amplitudes,
$$
\mathcal{M}_n^{\mathcal{N}_G=8}(\Phi_1,\Phi_2,\ldots,\Phi_n)  ~.
$$
When analogously expanded out in terms of the $\eta_{i,A}$'s, this
gives a sum of
all possible component amplitudes $M_n^{N^kMHV}$ dressed with
strings of $\eta_{i,A}$'s, where the capital letters run from 1 to 8.
The $SU(8)_R$ invariance dictates that only
amplitudes dressed with a string of $\eta_A$'s where each $A$ index
appears an equal amount of times can be non-vanishing.

In this on-shell formalism, the KLT-relations between maximally
supersymmetric supergravity and super Yang-Mills theories
can be formulated in an extremely compact way. As is perhaps now almost evident,
the $n$-point KLT-relations between $\mathcal{N}_G=8$ supergravity
superamplitudes $\mathcal{M}_n^{\mathcal{N}_G=8}$ and the product of
two $\mathcal{N}=4$ super Yang-Mills superamplitudes
$\widetilde{\mathcal{A}}_{n}^{\widetilde{\mathcal{N}}=4}$ and
$\mathcal{A}_{n}^{\mathcal{N}=4}$ can be wrapped into
\bea
\mathcal{M}_n^{\mathcal{N}_G=8} = \sum_{\gamma,\beta\in
S_{n-3}}
\widetilde{\mathcal{A}}_{n}^{\widetilde{\mathcal{N}}=4}(n-1,n,\gamma,1)
\mathcal{S}[\gamma|\beta]_{p_1}
\mathcal{A}_{n}^{\mathcal{N}=4}(1,\beta,n-1,n)~.~~~
\label{n844-KLT}
\eea
Here $\gamma$ and $\beta$ are again just permutations over the legs
$2,\ldots,n-2$ and the S-kernel is the same as defined in
(\ref{s-kernel}). $\mathcal{M}_n^{\mathcal{N}_G=8},
\widetilde{\mathcal{A}}_{n}^{\widetilde{\mathcal{N}}=4}$ and
$\mathcal{A}_{n}^{\mathcal{N}=4}$ are the superamplitudes of on-shell
superfields as defined above. This maximally supersymmetric
KLT-relation has been proven by
BCFW on-shell recursion relations by Feng and He \cite{FengHe}.
The superfield expansions on each side automatically yield all the
correct component relations when the $\eta$'s on the supergravity side
are correctly identified as the union of $\eta$'s of the two super Yang-Mills
theories. The explicit operator map of states between the two ${\mathcal N}=4$
super Yang-Mills theories and ${\mathcal N}_G=8$ supergravity has been worked out
in ref. \cite{Elvang1}.

Interestingly, the superamplitude formulation of the
maximally supersymmetric KLT-relation of eq.~(\ref{n844-KLT})
contains all information required to construct KLT-relations for theories with
less supersymmetry as well. Superamplitudes for supersymmetric
Yang-Mills theories with less than maximal supersymmetry were
introduced in \cite{Elvang2}.
Since $SU(8)_R\supset SU(4)_R\otimes SU(4)_R$, there is perfect matching between
$SU(4)_R$
indices $1,2,3,4$ of one $\widetilde{\mathcal{N}}=4$ super Yang-Mills amplitude
$\widetilde{\mathcal{A}}^{\widetilde{\mathcal{N}}=4}_n$ and $SU(4)_R$ indices
$5,6,7,8$ of the other amplitude $\mathcal{A}^{\mathcal{N}=4}_n$. The
product will label all states in the $\mathcal{N}_G=8$ supergravity theory
with $SU(8)_R$ indices $1,2,\ldots,8$. In order to get the component
KLT-relations, we need only to expand superamplitudes of eq.
(\ref{n844-KLT}) into their component amplitudes dressed with their
strings of $\eta_{i,A}$ and $\eta_{i,a},\eta_{i,b}$'s, where
$i=1,\ldots,n$, $A=1,\ldots,8$, $a=1,2,3,4$ and $b=5,6,7,8$. By
picking up the appropriate coefficients of $\eta$-strings on the
left and right sides of eq. (\ref{n844-KLT}), and taking care of signs when
exchanging Grassmann variables, we get the KLT-relations for component
amplitudes. For example, if we wish to get the KLT-relation for
a pure graviton MHV amplitude $M_n(h^-,h^-,h^+,\ldots,h^+)$, from the
$\mathcal{N}_G=8$ superfield expansion, we know that this amplitude is
just the coefficient of the string
$\prod_{A=1}^{8}\eta_{1,A}\prod_{A=1}^{8}\eta_{2,A}$ on the left
hand side of eq. (\ref{n844-KLT}). This string of $\eta$'s decomposes
into
$\prod_{a=1}^{4}\eta_{1,a}\prod_{a=1}^{4}\eta_{2,a}\prod_{b=5}^{8}\eta_{1,b}\prod_{b=5}^{8}\eta_{2,b}$,
whose coefficient on the right hand side is nothing but two pure-gluon MHV amplitudes
of proper helicities. Thus
the component KLT-relation for the pure graviton MHV amplitude is a
sum of a product of two pure gluon MHV amplitudes with the
S-kernel in-between and the corresponding permutations of the external legs.

The superamplitude version of KLT-relations is actually far more powerful.
In this maximally supersymmetric KLT-formulation we can consider the violation of
$SU(8)_R$ symmetry on the supergravity side, while each $SU(4)_R$ symmetry
of the super Yang-Mills side is kept intact. $\widetilde{\mathcal{A}}^{\widetilde{\mathcal{N}}=4}_n$
should be invariant under $SU(4)_R$, and thus the power $k$ of each
$\eta_a,a=1,2,3,4$ should be the same. Similarly, also
$\mathcal{A}^{\mathcal{N}=4}_n$ should be invariant under
$SU(4)_R$ and the power $k'$ of each $\eta_b,b=5,6,7,8$ should be
the same. Seen from the super Yang-Mills side of the KLT-relation, the power $k$
is not necessarily equal to $k'$, but the product of these two super Yang-Mills theory
amplitudes should produce a supergravity amplitude and hence possess $SU(8)_R$ symmetry.
Thus if $k\neq k'$, the resulting supergravity amplitude will obviously
violate $SU(8)_R$ symmetry and vanish \cite{FengHe,Elvang2}. We thereby get
vanishing identities of superamplitudes such as
\bea
 0 = \sum_{\gamma,\beta\in S_{n-3}}
\widetilde{A}_{n}^{N^kMHV}(n-1,n,\gamma,1)
\mathcal{S}[\gamma|\beta]_{p_1}
A_{n}^{N^{k'}MHV}(1,\beta,n-1,n)~,~~~\label{N8-vanishing}
\eea
where $k\neq k'$. As we describe in a little more detail below, this explains all the
vanishing identities originally found in \cite{BjerrumBohr6}, and gives the general
prescription for how to obtain these identities systematically. For the pure gauge theory case,
the analogous argument was given in ref. \cite{TyeZhang1}

We will present the non-maximal supersymmetric KLT-relations in the next section. These supersymmetric KLT-relations
  will also make the vanishing identities manifest in these cases.

\section{KLT-relations with less supersymmetry: the full map}


In this section, we derive the supersymmetric
KLT-relations for $\mathcal{N}_G<8$ supergravity theories, by
removing and integrating out components in the $\mathcal{N}_G=8$ formalism.

We begin with a brief review of the $\Phi$-$\Psi$ on-shell superfield
formalism for super Yang-Mills theory.
As is well-known, all states in a maximally supersymmetric theory are related under
the action of
supercharge generators $\widetilde{Q}_a$ and $Q^a$. This means that
starting from the highest helicity state $+h$ we can lower the
helicity by $1/2$ each time when acting with $\widetilde{Q}_a$ all the way
down to the lowest helicity state $-h$. Thus we can pack all states
into one superfield (\ref{Phi-N4}), and since
$\Phi^{\mathcal{N}=4}$ is CPT self-conjugate, this superfield is
already complete. But for super Yang-Mills theories with
$\mathcal{N}<4$, this is not true.
For example, for the $\mathcal{N}=2$ super Yang-Mills theory we have $\widetilde{Q}^a, a=1,2$.
Starting from the $+1$ helicity state we can at most lower the
helicity to $0$, and hence we cannot reach the CPT-conjugate state
of helicity $-1$.

Following \cite{Elvang2} we can always get one of the $\mathcal{N}<4$ superfields
$\Phi^{\mathcal{N}}$ from a truncation of the $\mathcal{N}=4$ superfield
by simply setting the unwanted $\eta$'s to zero, \textit{i.e.}
\bea
\Phi^{\mathcal{N}<4} =
\Phi^{\mathcal{N}=4}|_{\eta_{\mathcal{N}+1},\ldots,\eta_4
\rightarrow 0}~.~~~\label{Phi-truncation}
\eea
For example, by setting $\eta_4\to 0$ in (\ref{Phi-N4}), we get
\bea
\Phi^{\mathcal{N}=3} = g_+ + \eta_a f_+^a +
\frac{1}{2!}\eta_a\eta_b s^{ab} + \eta_1\eta_2\eta_3
f_-^{123}~,~~~\label{Phi-N3}
\eea
where $a,b=1,2,3$. This superfield contains one  plus-helicity gluon, three
plus-helicity fermions, three real scalars and one minus-helicity fermion. It is therefore
not complete and we should add the CPT-conjugate superfield.
 This additional superfield can be obtained by integrating out the unwanted $\eta$'s
in eq. (\ref{Phi-N4}). We therefore introduce another $\mathcal{N}<4$
superfield $\Psi^{\mathcal{N}}$ by
\bea
\Psi^{\mathcal{N}<4} = \int \prod_{a=\mathcal{N}+1}^4 d\eta_a
\Phi^{\mathcal{N}=4}~.~~~\label{Psi-truncation}
\eea
As an example, for $\mathcal{N}=3$, we have
\bea
\Psi^{\mathcal{N}=3} =f_+^{(4)} -\eta_a s^{a(4)} +
\frac{1}{2!}\eta_a\eta_b f_-^{ab(4)} -\eta_1\eta_2\eta_3
g_-^{123(4)}~,~~~\label{Psi-N3}
\eea
where $a,b=1,2,3$. The index $4$ is placed in parenthesis because
the corresponding Grassmann parameter $\eta_4$ has been
integrated out. We have only indices $1,2,3$ for the
$SU(3)_R$ symmetry in $\mathcal{N}=3$ super Yang-Mills theory, and the index $4$ is what we will call a {\sl hidden }
index from $SU(4)_R$. Although it has nothing to do with the $SU(3)_R$ symmetry,
we keep it for reasons that will be
explained in the next section. Using the combined  $\Phi$-$\Psi$ formalism, all
states now have their CPT-conjugate partners, and it is thus complete. The
$\mathcal{N}<4$ superamplitudes can now directly be obtained from the
$\mathcal{N}=4$ superamplitudes. Suppose the $i_1<i_2<
\ldots<i_m$ external legs are in the $\Psi$ superfield
representation, while the $j_1<j_2<\ldots<j_l$ external legs are in the $\Phi$ representation and
$m+l=n$. The $\mathcal N<4$ superamplitude is then 
\bea
\mathcal{A}_{n,i_1\ldots i_m}^{\mathcal{N}<4} = \left[ \int
\prod_{a_1=\mathcal{N}+1}^4 d\eta_{i_1,a_1}\cdots
\prod_{a_m=\mathcal{N}+1}^4 d\eta_{i_m,a_m}
\mathcal{A}_n^{\mathcal{N}=4}(\Phi_1,\Phi_2,\ldots,\Phi_n)
\right]_{\eta_{\mathcal{N}+1},\ldots,\eta_4 \rightarrow 0}~.~~~
\label{AN<4}
\eea

\begin{figure}
\center
  \includegraphics[width=6in]{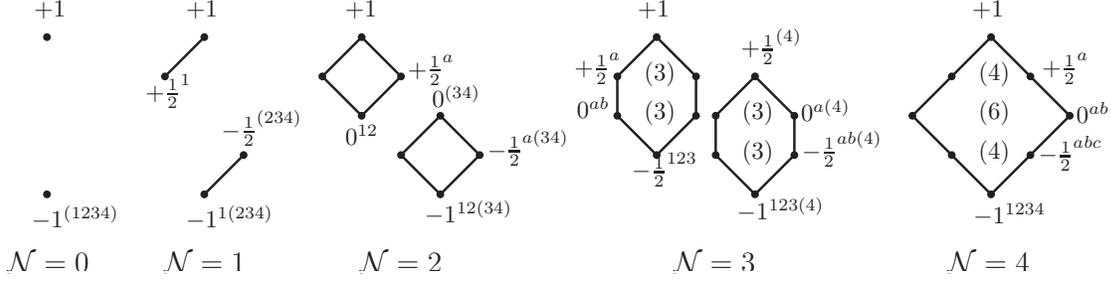}\\
  \caption{Diamond diagrams for superfields of super Yang-Mills theories with increasing amount
of supersymmetry. The $SU(\mathcal{N})_R$ indices $a,b,c$ are labeled
  as superscripts, where $a<b<c$ with $a,b,c=1,2,\ldots,\mathcal{N}$.
  The hidden indices, which refer to where they originated from in the maximally supersymmetric
multiplet are indicated in parentheses. The numbers inside the diamonds show
  the number of corresponding states on each horizontal line.}\label{diamond}
\end{figure}

In order to keep track of fields in a systematic and graphical manner, we
introduce a diagrammatic notation which we have found
useful. The idea is to express the superfields in terms of (generalized)
``diamond diagrams", which keep track of helicities and $R$-symmetry
quantum numbers. A diamond is a set of
on-shell component fields which are all related by supersymmetric
transformations. A diamond may not be a full supermultiplet since it may not
be CPT complete. We illustrate this in figure \ref{diamond} for all four
 super Yang-Mills cases.
We use solid lines to connect all states that are related under operation by
supercharges. This means that we can reach all states within one diamond
by applying $\widetilde{Q}_a$ or $Q^a$  a sufficient number of times on any arbitrary
initial state. In contrast, there is no way to reach states in a different
diamond by a similar procedure. The diamonds
without hidden indices represent $\Phi$ superfields while those
with hidden indices represent $\Psi$ superfields, the CPT
conjugates of $\Phi$ superfields. Superamplitudes of these
superfields should be invariant under their $SU(\mathcal{N})_R$
transformation. This puts  constraints on the corresponding
$SU(\mathcal{N})_R$ indices $a,b,c$, but clearly not on the hidden indices
inside parentheses.

The convenience of using this diamond representation is that we
straightforwardly can obtain
all states on the supergravity side from the product of two super Yang-Mills theories
while we explicitly keep track of both the $SU(\mathcal{N})_R$ and the hidden
indices. It is also easy to count the number of states on the
supergravity side.  We can construct diamonds for superfields of all the different supergravity
theories. For maximal $\mathcal{N}_G=8$ supergravity we evidently need only one
diamond, which contains all states from helicity $+2$ to helicity $-2$.
Analogously to the situation on the gauge theory side,
for $\mathcal{N}_G<8$ we need more than one diamond to
express a complete set of states in the supergravity theory.

\subsection{The equivalence of $\mathcal{N}=3$ and $\mathcal{N}=4$ super Yang-Mills theory}

Before proceeding to the $\mathcal{N}_G<8$ super-KLT relations, we want to briefly
dwell on the equivalence of $\mathcal{N}=3$ and
$\mathcal{N}=4$ super Yang-Mills theories. This is textbook material,
but it is instructive to see it in the light of our diamond representation.
We find it directly from the
diamonds in figure \ref{diamond}: for the $\mathcal{N}=3$
superfields, the two diamonds combined will contribute with one gluon, four
fermions and six real scalars, which is exactly the same as the field
content of $\mathcal{N}=4$ super Yang-Mills theory. We can recover
$\eta_4$ of $\Psi^{\mathcal{N}=3}$ and combine the two superfields
$\Phi^{\mathcal{N}=3}$ and $\Psi^{\mathcal{N}=3}$ in the following
way
\bea
\Phi^{\mathcal{N}=3}+\eta_4\Psi^{\mathcal{N}=3}~.~~~
\eea
Using results of eqs. (\ref{Phi-N3}) and (\ref{Psi-N3}) we immediately see that
\bea
\Phi^{\mathcal{N}=3}+\eta_4\Psi^{\mathcal{N}=3}=\Phi^{\mathcal{N}=4}~,~~~
\eea
which means that the two $\mathcal{N}=3$ superfields are nothing
but a rewriting of the $\mathcal{N}=4$ superfield. For the
$\mathcal{N} \neq 3$ superfields there is no way of combining the
$\Phi$ and $\Psi$ superfields into one of larger $\mathcal{N}$. In
the following discussion, we will simply treat the $\mathcal{N}=3$
theory as equivalent to the $\mathcal{N}=4$ theory. In supergravity
there is a completely equivalent phenomenon associated with the
$\mathcal{N}_G=7$ superfields which can be combined into that of
$\mathcal{N}_G=8$,
\bea
\Phi^{\mathcal{N}_G=7}+\eta_8\Psi^{\mathcal{N}_G=7}=\Phi^{\mathcal{N}_G=8}~,~~~
\eea
so that also the  $\mathcal{N}_G=7$ and  $\mathcal{N}_G=8$ supergravity theories
are equivalent. Actually, since we intend to construct the supergravity theories
from gauge theories
through super-KLT relations, we will only assume the
equivalence of the $\mathcal{N}=3$ and $\mathcal{N}=4$ gauge theories. The
equivalence of $\mathcal{N}_G=7$ and $\mathcal{N}_G=8$ supergravity theories is then
an obvious consequence of the former equivalence.

\subsection{Diamond diagrams and the $\mathcal{N}_G<8$ KLT-relations}

In order to obtain $\mathcal{N}<4$ superfields from the
$\mathcal{N}=4$ superfield, we set to zero or
integrate out the unwanted $\eta_a$'s, as prescribed in eq.~\eqref{Phi-truncation} and \eqref{Psi-truncation}. Their resulting
superfields can be expressed graphically in terms of our
diamond diagrams, as shown in figure
\ref{diamond}. Because of the super-KLT relation between two $\mathcal{N}=4$
gauge theories to $\mathcal{N}_G=8$ supergravity, it is possible to
write not only the superfields of supergravity theories with smaller $\mathcal{N}_G$ in terms of
combinations of two gauge superfields, but also the supergravity
superamplitude in those variables.

The identification of the expressions for the $\mathcal{M}^{\mathcal{N}_G \leq 8}_n$
superamplitudes will be done in terms of its embedding in $\mathcal{N}_G=8$
  supergravity and the use of the canonical splitting of the $SU(8)_R$ indices into the two
subsets $1,2,3,4$ and $5,6,7,8$,  for the
  two super Yang-Mills superamplitudes $\widetilde{\mathcal A}^{\widetilde{\mathcal N}}_n$ and $\mathcal
  A^{\mathcal N}_n$, respectively.  So in general, we classify the external states of the supergravity superamplitudes
into one of four representations, $(\widetilde \Phi, \Phi)$, $(\widetilde \Psi, \Phi)$, $(\widetilde \Phi,
\Psi)$, $(\widetilde \Psi, \Psi)$, which will be explained in greater details below (here $\widetilde{\Phi}$ and $\widetilde{\Psi}$ is
just shorthand for $\Phi^{\widetilde{\mathcal{N}}}$ and $\Psi^{\widetilde{\mathcal{N}}}$, respectively). In particular, if $\widetilde{\mathcal
  N}=4$, then $\widetilde \Phi=\widetilde \Psi$ and the number of
representations is reduced by a factor of $2$.

By taking the KLT-product between two
arbitrary superamplitudes we thereby get
\bea
 &&\sum_{\gamma,\beta\in S_{n-3}}
\widetilde{\mathcal{A}}_{n,\tilde i_1 ,\ldots \tilde
  i_{\widetilde m}}^{\widetilde{\mathcal{N}}\leq
4}(n-1,n,\gamma,1) \mathcal{S}[\gamma|\beta]_{p_1}
\mathcal{A}_{n, i_1\ldots i_m}^{\mathcal{N}\leq 4}(1,\beta,n-1,n) \nonumber \\
&&=\sum_{\gamma,\beta\in S_{n-3}}  \left[ \int
\prod_{\tilde a_1=\widetilde{\mathcal{N}}+1}^4 d\eta_{\tilde
  i_1,\tilde a_1}\cdots
\hspace{-0.3cm} \prod_{\tilde a_{\widetilde m}=\widetilde{\mathcal{N}}+1}^4 d\eta_{\tilde
  i_{\widetilde m},\tilde a_{\widetilde m}}
\widetilde{\mathcal{A}}_n^{\widetilde{\mathcal{N}}=4}(n-1,n,\gamma,1)
\right]_{\eta_{\widetilde{\mathcal{N}}+1},\ldots,\eta_4 \rightarrow 0} \nonumber \\
&&\hspace{1cm}\times \mathcal{S}[\gamma|\beta]_{p_1} \times \left[
\int \prod_{a_1=\mathcal{N}+5}^8 d\eta_{i_1,a_1}\cdots
\hspace{-0.3cm}\prod_{a_m=\mathcal{N}+5}^8 d\eta_{i_m,a_m}
\mathcal{A}_n^{\mathcal{N}=4}(1,\beta,n-1,n) \right]_{\eta_{\mathcal{N}+5},\ldots,\eta_8 \rightarrow 0} \nonumber \\
&&= \left[ \int \prod_{\tilde a_1=\widetilde{\mathcal{N}}+1}^4 d\eta_{\tilde
  i_1,\tilde a_1}\cdots \hspace{-0.3cm}
\prod_{\tilde a_{\widetilde m}=\widetilde{\mathcal{N}}+1}^4 d\eta_{\tilde
  i_{\widetilde m},\tilde a_{\widetilde m}}\prod_{a_1=\mathcal{N}+5}^8 d\eta_{i_1,a_1}\cdots
\hspace{-0.3cm}\prod_{a_m=\mathcal{N}+5}^8 d\eta_{i_m,a_m}\right. \nonumber \\
&&\hspace{1cm}\left. \times \sum_{\gamma,\beta\in S_{n-3}}
\widetilde{\mathcal{A}}_{n}^{\widetilde{\mathcal{N}}=4}(n-1,n,\gamma,1)
\mathcal{S}[\gamma|\beta]_{p_1}
\mathcal{A}_{n}^{\mathcal{N}=4}(1,\beta,n-1,n)\right]_{\substack{\eta_{\widetilde{\mathcal{N}}+1},\ldots,\eta_4  \rightarrow  0 \\
\eta_{\mathcal{N}+5},\ldots,\eta_8  \rightarrow  0}} \nonumber \\
&&=\left[ \int\prod_{\tilde a_1=\widetilde{\mathcal{N}}+1}^4 d\eta_{\tilde
  i_1,\tilde a_1}\cdots \hspace{-0.3cm}
\prod_{\tilde a_{\widetilde m}=\widetilde{\mathcal{N}}+1}^4 d\eta_{\tilde
  i_{\widetilde m},\tilde a_{\widetilde m}} \right. \nonumber \\
&&\hspace{2.5cm} \left. \times \prod_{a_1=\mathcal{N}+5}^8 d\eta_{i_1,a_1}\cdots
\hspace{-0.3cm}\prod_{a_m=\mathcal{N}+5}^8 d\eta_{i_m,a_m}
\mathcal{M}_n^{\mathcal{N}_G=8}(\Phi_1,\Phi_2,\ldots,\Phi_n)\right]_{\substack{\eta_{\widetilde{\mathcal{N}}+1},\ldots,\eta_4  \rightarrow  0 \\
\eta_{\mathcal{N}+5},\ldots,\eta_8  \rightarrow  0}} \nonumber \\
&&\equiv \mathcal{M}_{n,(\tilde i_1 ,\ldots, \tilde
  i_{\widetilde m});  (i_1,\ldots , i_m)}^{\mathcal{N}_G\leq
8}~,~~~\label{KLT-product}
\eea
where the subscripts $(\tilde i_1 ,\ldots, \tilde
  i_{\widetilde m})$ and $ (i_1,\ldots, i_m)$ label the external legs given by  $\widetilde \Psi$ and $\Psi$ fields,
    respectively. $\widetilde m \leq n$ and $m \leq n$. In the second to last
  step, we used the $\mathcal N_G=8$ super KLT-relation. $\mathcal M ^{\mathcal{N}_G\leq 8}_n$
in the last line is the superamplitude for  $\mathcal{N}_G\leq
8$ supergravity, obtained from $\mathcal N_G=8$ by setting to zero or
integrating out unwanted $\eta$'s. To see this explicitly, we present
the four possible superfields for an external leg $k$.

\begin{itemize}
\item  $(\widetilde \Phi,\Phi)$: if $k \not \in (\tilde i_1,\ldots,\tilde i_{\widetilde m})$ and
  $k\not \in (i_1,\ldots,i_m)$, we set all
$\eta_{k,\widetilde{\mathcal{N}}+1},\ldots,\eta_{k,4}$ and
$\eta_{k,\mathcal{N}+5},\ldots,\eta_{k,8}$ to zero, and the
resulting superfield is
\bea
\Phi_k^{\mathcal{N}_G=\widetilde{\mathcal{N}}+\mathcal{N}}
=\Phi^{\mathcal{N}_G=8}_k|_{\eta_{k,\widetilde{\mathcal{N}}+1},\ldots,\eta_{k,4};\eta_{k,\mathcal{N}+5},\ldots,\eta_{k,8}\to
0}~.~~~
\eea
\item $(\widetilde \Psi,\Psi)$: if $k  \in (\tilde i_1,\ldots,\tilde i_{\widetilde m})$ and
  $k \in (i_1,\ldots,i_m)$, we get
another superfield
\bea
\Psi_k^{\mathcal{N}_G=\widetilde{\mathcal{N}}+\mathcal{N}}=
\int\prod_{a=\widetilde{\mathcal{N}}+1}^{4}d\eta_{k,a}
\prod_{b=\mathcal{N}+5}^{8}d\eta_{k,b}\Phi^{\mathcal{N}_G=8}_k~.~~~
\eea
These two superfields combine to form a full $SU(\mathcal{N}_G)$
supergravity multiplet.
\item $(\widetilde \Psi, \Phi)$: if $k  \in (\tilde i_1,\ldots,\tilde i_{\widetilde m})$ and
  $k \not \in (i_1,\ldots,i_m)$, we have
\bea
\Theta_k^{\mathcal{N}_G=\widetilde{\mathcal{N}}+\mathcal{N}}=
\int\prod_{a=\widetilde{\mathcal{N}}+1}^{4}d\eta_{k,a}
\Phi^{\mathcal{N}_G=8}_k|_{\eta_{k,\mathcal{N}+5},\ldots,\eta_{k,8}\to
0}~.~~~
\eea
\item
$(\widetilde \Phi, \Psi)$: $k  \not \in (\tilde i_1,\ldots,\tilde i_{\widetilde m})$ and
  $k \in (i_1,\ldots,i_m)$,
\bea
\Gamma_k^{\mathcal{N}_G=\widetilde{\mathcal{N}}+\mathcal{N}}=
\int\prod_{b=\mathcal{N}+5}^{8}d\eta_{k,b}
\Phi^{\mathcal{N}_G=8}_k|_{\eta_{k,\widetilde{\mathcal{N}}+1},\ldots,\eta_{k,4}\to
0}~.~~~
\eea
The latter two superfields combine to form an $SU(\mathcal{N}_G)$
matter supermultiplet, if $\widetilde{\mathcal N}<3$ and $\mathcal N<3$.
\end{itemize}

Thus 
$ \mathcal{M}_{n,(\tilde i_1 ,\ldots, \tilde i_{\widetilde m});  (i_1,\ldots, i_m)}^{\mathcal{N}_G\leq8}$ in eq.
(\ref{KLT-product}) is the $\mathcal{N}_G\leq 8$ supergravity amplitude based on all four superfields
$\Phi^{\mathcal{N}_G},\Psi^{\mathcal{N}_G},\Theta^{\mathcal{N}_G}$
and $\Gamma^{\mathcal{N}_G}$. This completes the
formal construction of all $\mathcal N_G<8$ super KLT-relations.

\begin{figure}
\center
  \includegraphics[width=5in]{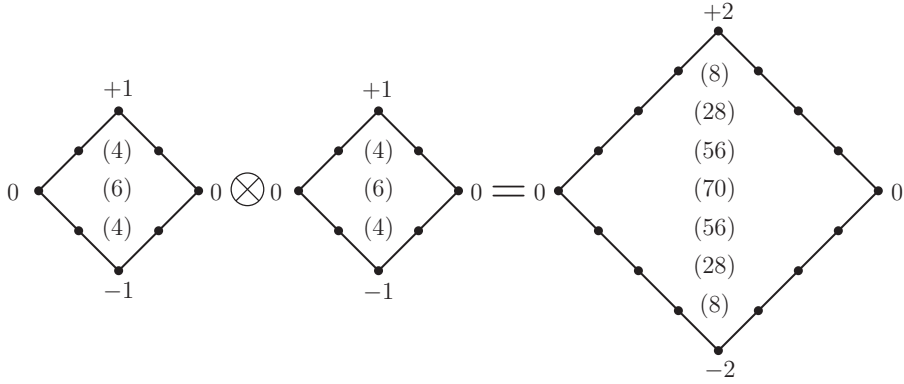}\\
  \caption{Diamond diagrams that demonstrates the matching of states in
(supergravity)$_{\mathcal{N}_G=8}$ = (super Yang-Mills)$_{\widetilde{\mathcal{N}}=4}\otimes$
(super Yang-Mills)$_{\mathcal{N}=4}$.
 The numbers inside the diamonds indicate the number of states on each line,
and the number next to the dots indicate the helicities.
Only the highest, lowest and zero helicities have been labeled explicitly.}\label{klt844}
\end{figure}

Since the states of the supergravity theories arise from the tensor product
between two super Yang-Mills states, we can also label the superfield of
the pertinent supergravity theory by the tensor
product of two gauge theory superfields through our diamond diagrams.
As an example, we illustrate this by the
diamond diagram of the $\mathcal{N}_G=8$ super KLT-relation. This is shown in
figure \ref{klt844}.
{}From this we can very easily establish how all states of the $\mathcal{N}_G=8$ superfield
arise from the tensor product;
the one graviton state $h^{\pm}$ comes from $$(+1)\otimes (+1)~~
\mbox{and}~~ (-1)\otimes (-1)~,~$$
the  $8\times 2$  gravitino states $\psi^{\pm}$  from
$$(+1/2)^4\otimes (+1)~,~ (+1)\otimes(+1/2)^4~~\mbox{and}~~ (-1/2)^4\otimes
(-1)~,~(-1)\otimes(-1/2)^4,$$
the  $28\times 2$ vector states $v^{\pm}$ arise through
$$(0)^6\otimes
(+1)~,~(+1/2)^4\otimes(+1/2)^4~,~(+1)\otimes(0)^6~~\mbox{and}~~(0)^6\otimes
(-1)~,~(-1/2)^4\otimes(-1/2)^4~,~(-1)\otimes(0)^6~,~~$$
the  $56\times 2$ spin-1/2 fermions
$\chi^{\pm}$ are built out of $$(-1/2)^4\otimes (+1)~,~(0)^6\otimes
(+1/2)^4~,~(+1/2)^4\otimes (0)^6~,~(+1)\otimes (-1/2)^4~,~~~$$ $$
(+1/2)^4\otimes (-1)~,~(0)^6\otimes (-1/2)^4~,~(-1/2)^4\otimes
(0)^6~,~(-1)\otimes (+1/2)^4~,~~$$
and finally the 70 scalars
$$(-1)\otimes
(+1)~,~(-1/2)^4\otimes (+1/2)^4~,~(0)^4\otimes
(0)^4~,~(+1/2)^4\otimes(-1/2)^4~,~(+1)\otimes(-1)~,~~$$ where the
superscripts denote the degeneracies of the states. This is exactly the
field content of the $\mathcal{N}_G=8$ supergravity theory. The $SU(8)_R$ indices
can be recovered by combining two sets of $SU(4)_R$ indices. This is
shown in figure \ref{diamond}.

In a similar manner we will now obtain all possible $\mathcal{N}_G<8$
supergravity theories that can be constructed from KLT products.
We will explicitly work out their field content and the corresponding
diamond diagrams. This gives
us the complete table of supergravity theories obtained from tensor products
of super Yang-Mills theories in four dimensions. We have for convenience summarized
our results in table \ref{table-sugra}.

\begin{table}
\center
\begin{tabular}{|c|c|l|}
  \hline
  $\mathcal{N}_G$ & $\widetilde{\mathcal{N}}\otimes\mathcal{N}$ & Description \\
  \hline
  8 & $4\otimes 4$ & Maximal $\mathcal{N}_G=8$ Supergravity \\
  \hline
  7 & $4\otimes 3$ & Maximal $\mathcal{N}_G=8$ Supergravity \\
  \hline
  6 & $4\otimes 2$ & Minimal $\mathcal{N}_G=6$ Supergravity with $SU(6)$ supergravity multiplet\\
  \hline
  6 & $3\otimes 3$ & Maximal $\mathcal{N}_G=8$ Supergravity\\
  \hline
  5 & $4\otimes 1$ & Minimal $\mathcal{N}_G=5$ Supergravity with $SU(5)$ supergravity multiplet\\
  \hline
  5 & $3\otimes 2$ & Minimal $\mathcal{N}_G=6$ Supergravity with $SU(6)$ supergravity multiplet\\
  \hline
  4 & $4\otimes 0$ & Minimal $\mathcal{N}_G=4$ Supergravity with $SU(4)$ supergravity multiplet\\
  \hline
  4 & $3\otimes 1$ & Minimal $\mathcal{N}_G=5$ Supergravity with $SU(5)$ supergravity multiplet\\
  \hline
  4 & $2\otimes 2$ & $\mathcal{N}_G=4$ Supergravity multiplet coupled to vector multiplet\\
  \hline
  3 & $3\otimes 0$ & Minimal $\mathcal{N}_G=4$ Supergravity with $SU(4)$ supergravity multiplet\\
  \hline
  3 & $2\otimes 1$ & $\mathcal{N}_G=3$ Supergravity multiplet coupled to vector multiplet \\
  \hline
  2 & $2\otimes 0$ & $\mathcal{N}_G=2$ Supergravity multiplet coupled to vector multiplet \\
  \hline
  2 & $1\otimes 1$ & $\mathcal{N}_G=2$ Supergravity multiplet coupled to hypermultiplet \\
  \hline
  1 & $1\otimes 0$ & $\mathcal{N}_G=1$ Supergravity multiplet coupled to chiral multiplet \\
  \hline
  0 & $0\otimes 0$ & Einstein gravity coupled to two scalars \\
  \hline
\end{tabular}
\caption{Full list of all possible supergravity theories that can be constructed
from KLT-relations of minimal super Yang-Mills theories with varying degree of supersymmetry.} \label{table-sugra}
\end{table}
We can classify all of these theories into three categories. Category
I consists of maximal $\mathcal{N}_G=8$ supergravity, its equivalent
$\mathcal{N}_G=7$ supergravity theory, and the
$(\widetilde{\mathcal{N}}=3)\otimes(\mathcal{N}=3)$ theory. Since
their field contents are all the same, one must consider them as describing the
same theory just encoded in slightly different ways. The super KLT-relations for
these theories are equivalent to eq. (\ref{n844-KLT}).

Category II contains all
minimal supergravity theories with $4\leq\mathcal{N}_G<8$. These theories
contain only the minimal supergravity multiplet 
  (the one-graviton supermultiplet). They are given in terms of two diamonds (which arise from the two
superfields  $\Phi$ and $\Psi$) to describe the complete set of states. These
theories all arise from the KLT product
$(\widetilde{\mathcal{N}}=4)\otimes(\mathcal{N}\leq 2)$ (or $(\widetilde{\mathcal{N}}=3)\otimes(\mathcal{N}\leq 2)$ due to the equivalence
between $\widetilde{\mathcal{N}}=3$ and $\widetilde{\mathcal{N}}=4$). The corresponding super KLT-relations can be expressed as
\bea
&&\mathcal{M}_n^{\mathcal{N}_G=
4+\mathcal{N}}(\Phi^{\mathcal{N}_G}_{i_1, \ldots, i_{m_1}},\Psi^{\mathcal{N}_G}_{j_1, \ldots, j_{m_2}})=\nonumber\\
&&~~\sum_{\gamma,\beta\in S_{n-3}}
\widetilde{\mathcal{A}}_n^{\widetilde{\mathcal{N}}=4}
(\Phi^{\widetilde{\mathcal{N}}=4}_{1,\ldots,n})\times\mathcal{S}[\gamma|\beta]_{p_1} \times
\mathcal{A}_n^{\mathcal{N}\leq 2}(\Phi^{\mathcal{N}\leq 2}_{i_1, \ldots, i_{m_1}},\Psi^{\mathcal{N}\leq 2}_{j_1, \ldots, j_{m_2}})~,~~~
\eea
where the indices $(i_1, \ldots, i_{m_1})$ and $(j_1, \ldots, j_{m_2})$
denote the legs of the corresponding superfields and $m_1 + m_2 = n$.

Category III includes the remaining theories: they describe minimal
supergravity coupled to a variety of matter multiplets, and requires four diamonds to describe the
full CPT-complete state space.
This means that we have four kinds of superfields: $\Phi$-$\Psi$,
and $\Theta$-$\Gamma$. The super KLT-relation can be compactly expressed as
\bea
&&\mathcal{M}_n^{\mathcal{N}_G=\widetilde{\mathcal{N}}+\mathcal{N}}(\Phi^{\mathcal{N}_G}_{i_1,\ldots,i_{m_1}},
\Psi^{\mathcal{N}_G}_{j_1,\ldots,j_{m_2}},\Theta^{\mathcal{N}_G}_{k_1,\ldots,k_{m_3}},\Gamma^{\mathcal{N}_G}_{l_1,\ldots,l_{m_3}})=\nonumber\\
&&~~~~~~~~~~~~~~~\sum_{\gamma,\beta\in S_{n-3}}
\widetilde{\mathcal{A}}_n^{\widetilde{\mathcal{N}}\leq 2}(\Phi^{\widetilde{\mathcal{N}}\leq 2}_{i_1,\ldots,i_{m_1},l_1,\ldots,l_{m_3}},
\Psi^{\widetilde{\mathcal{N}}\leq 2}_{j_1,\ldots,j_{m_2},k_1,\ldots,k_{m_3}})\nonumber\\
&&~~~~~~~~~~~~~~~~~~~~~~~~~~\times \mathcal{S}[\gamma|\beta]_{p_1}
\times
\mathcal{A}_n^{\mathcal{N}\leq 2}(\Phi^{\mathcal{N}\leq 2}_{i_1,\ldots,i_{m_1},k_1,\ldots,k_{m_3}},
\Psi^{\mathcal{N}\leq 2}_{j_1,\ldots,j_{m_2},l_1,\ldots,l_{m_3}})~,~~~
\label{KLT_Cat_III}
\eea
where again
$(i_1,\ldots,i_{m_1}),(j_1,\ldots,j_{m_2}),(k_1,\ldots,k_{m_3})$ and
$(l_1,\ldots,l_{m_3})$ label legs of the corresponding superfields and $m_1 + m_2 + 2m_3 =n$. Note that
the number of $\Theta$ superfields must match the number of
$\Gamma$ superfields, otherwise the $SU(\mathcal{N}_G)_R$
symmetry will be violated.

It is interesting to note that there are no 
  KLT-maps (from minimal super Yang-Mills theories) that provide
just minimal supergravity multiplets for $\mathcal{N}_G < 4$: in all
cases we get supergravity coupled to matter multiplets. However, if we want to project out
the subset of KLT-relations that would appear in these minimal supergravity theories we need to
restrict the sum over indices on the right hand side of the
KLT-relations. We can achieve such a projection by
making the restriction $m=\widetilde m$ and $\tilde i_j = i_j$ for $j=1,\ldots,m$
in eq.~\eqref{KLT-product}. 

There are also vanishing identities for the $\mathcal{N}<4$ super Yang-Mills
amplitudes that follow from these $\mathcal{N}_G<8$ super KLT-relations much like
the ones in eq. (\ref{N8-vanishing}) for the maximally supersymmetric case.
We will later see how to explain the vanishing identities from the point of view
of less than maximal supersymmetry.

We will now go through all the different cases summarised in table \ref{table-sugra} using the diamond diagrams.
The explicit expressions for the superfields can be found in Appendix \ref{app1}.

\subsubsection{Diamond diagrams for the $\mathcal{N}_G=7$ theory}
%
\begin{figure}[h]
\center
  \includegraphics[width=5.5in]{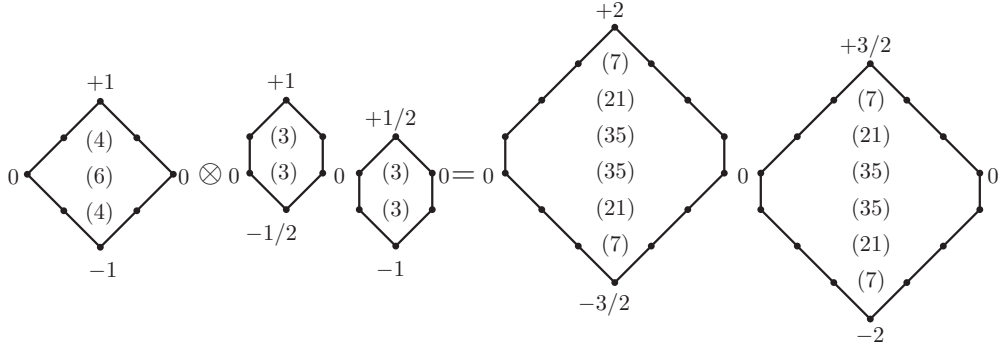}\\
  \caption{Diamond diagram for (supergravity)$_{\mathcal{N}_G=7}$ =
(super Yang-Mills)$_{\widetilde{\mathcal{N}}=4}\otimes$(super Yang-Mills)$_{\mathcal{N}=3}$.
The two diamonds
represent the $\Phi^{\mathcal{N}_G=7}$ and $\Psi^{\mathcal{N}_G=7}$ superfields, respectively.
Note that there is a hidden $SU(8)_R$ index in the $\Psi$ field,
for instance, the negative helicity graviton
state is $-2^{1234567(8)}$.}\label{klt734}
\end{figure}
The KLT-product between $\widetilde{\mathcal{N}}=4$ and $\mathcal{N}=3$ super Yang-Mills superamplitudes
maps exactly to $\mathcal{N}_G=7$ supergravity. This is illustrated by the tensor product between the
superfields represented by diamonds in figure \ref{klt734}. The $SU(\mathcal{N})_R$ indices for
component fields of the super Yang-Mills theories have already been shown in figure \ref{diamond}.
The $SU(\mathcal{N}_G)_R$ indices  come from combining
these two sets. The index ``8'' is now a hidden index in the
$\mathcal{N}=3$ diamond, and thus it is also a hidden index for the
$\Psi^{\mathcal N_G=7}$ superfield, which is shown as the second
diamond on the right hand side of figure \ref{klt734}. From this it is easy to see that
\bea
\Phi^{\mathcal{N}_G=8}=\Phi^{\mathcal{N}_G=7}+\eta_8\Psi^{\mathcal{N}_G=7}~,~~~
\eea
which again displays the fact that $\mathcal{N}_G=7$ supergravity is just a rewriting of
$\mathcal{N}_G=8$ supergravity. Likewise, the content of the super
KLT-relation for $\mathcal{N}_G=7$ is the same as the $\mathcal{N}_G=8$ version in figure \ref{klt844},
just encoded in a slightly different way.

\subsubsection{Diamond diagrams for the $\mathcal{N}_G=6$ theories}

%
\begin{figure}[h]
\center
  \includegraphics[width=5in]{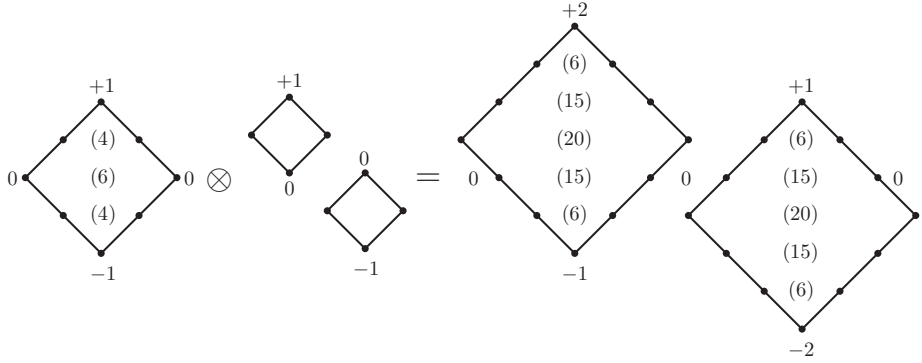}\\
  \caption{Diamond diagram for (supergravity)$_{\mathcal{N}_G=6}$ =
(super Yang-Mills)$_{\widetilde{\mathcal{N}}=4}\otimes$
(super Yang-Mills)$_{\mathcal{N}=2}$. The two diamonds represent the
$\Phi^{\mathcal{N}_G=6}$ and $\Psi^{\mathcal{N}_G=6}$ superfields,
respectively. There are two hidden indices $(78)$ for the $\Psi$ field.
}\label{klt624}
\end{figure}

The case of $(\widetilde{\mathcal{N}}=4)\otimes (\mathcal{N}=2)$ is rather
straightforward. There is an exact correspondence between minimal $\mathcal{N}_G=6$ supergravity
superamplitudes and the KLT-product between $\widetilde{\mathcal{N}}=4$
and $\mathcal{N}=2$ super Yang-Mills superamplitudes. This gravity theory consists
of 1 graviton $h_{\pm}$, 6 gravitinos $\psi_{\pm}$, 16 vectors $v_{\pm}$, 26 spin-1/2 fermions
$\chi_{\pm}$ and 30 scalars. This is illustrated in figure \ref{klt624}.

\begin{figure}[h]
\center
  \includegraphics[width=6in]{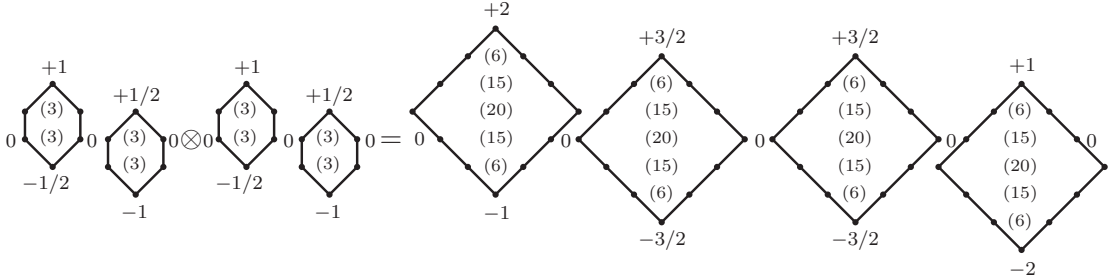}\\
  \caption{Diamond diagram for (supergravity)$_{\mathcal{N}_G=6}$ =
(super Yang-Mills)$_{\widetilde{\mathcal{N}}=3}\otimes$
(super Yang-Mills)$_{\mathcal{N}=3}$. The four diamonds correspond to the four superfields
$\Phi^{\mathcal{N}_G=6}$, $\Theta^{\mathcal{N}_G=6}$, $\Gamma^{\mathcal{N}_G=6}$
and $\Psi^{\mathcal{N}_G=6}$. The hidden indices are
$(4)$ for $\Theta$, $(8)$ for $\Gamma$ and $(48)$ for
$\Psi$ superfield.}\label{klt633}
\end{figure}

The case of $(\widetilde{\mathcal{N}}=3)\otimes (\mathcal{N}=3)$ is more interesting. There will be four diamonds,
as shown in figure \ref{klt633}, from where it is straightforward to see that
\bea
\Phi^{\mathcal{N}_G=8}=\Phi^{\mathcal{N}_G=6}+\eta_4\Theta^{\mathcal{N}_G=6}+
\eta_8\Gamma^{\mathcal{N}_G=6}+
\eta_4\eta_8\Psi^{\mathcal{N}_G=6}~.~~~
\eea
This is just a rewriting of maximal $\mathcal{N}_G=8$
supergravity. This is expected since $\mathcal{N}=3$ super Yang-Mills is just a rewriting
of $\mathcal{N}=4$ super Yang-Mills. However, using the aforementioned restriction
$m=\widetilde m$ and $\tilde i_j = i_j$ for $j=1,\ldots,m$
in eq.~\eqref{KLT-product}, we can project out the $\mathcal{N}_G=6$ superamplitudes from the
$(\widetilde{\mathcal{N}}=3)\otimes (\mathcal{N}=3)$ product if desired.

\subsubsection{Diamond diagrams for the $\mathcal{N}_G=5$ theories}

%
\begin{figure}[h]
\center
  \includegraphics[width=4.5in]{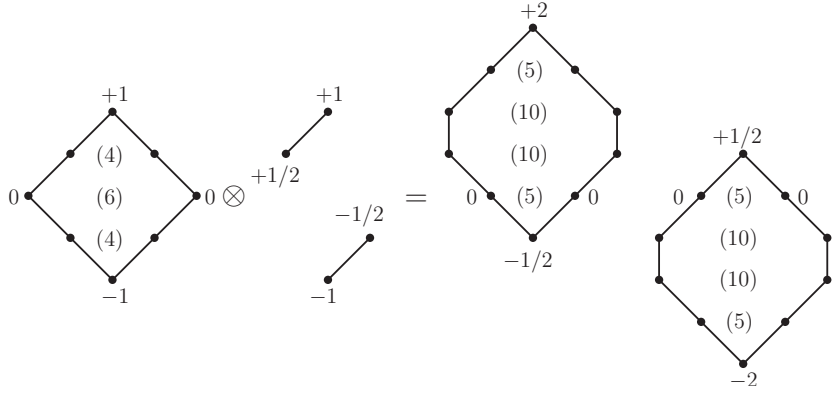}\\
  \caption{Diamond diagram for (supergravity)$_{\mathcal{N}_G=5}$
= (super Yang-Mills)$_{\widetilde{\mathcal{N}}=4}\otimes$
(super Yang-Mills)$_{\mathcal{N}=1}$. The two diamonds represent
the $\Phi^{\mathcal{N}_G=5}$ and $\Psi^{\mathcal{N}_G=5}$ superfields.
The are hidden indices $(678)$ for the $\Psi$ field.}\label{klt514}
\end{figure}

There is again two cases to consider. The first one is from the product between $\widetilde{\mathcal{N}}=4$ and
$\mathcal{N}=1$ super Yang-Mills theory, see figure \ref{klt514}. There is
one $(+2,+{3\over 2}^5,+1^{10},+{1\over 2}^{10},0^5,-{1\over
2}^1)$ supergravity diamond plus its CPT conjugate, and a total of
1 graviton $h_{\pm}$, 5 gravitinos $\psi_{\pm}$, 10 vectors $v_{\pm}$, 11 spin-1/2
fermions $\chi_{\pm}$ and 10 scalars. This is minimal $\mathcal{N}_G=5$ supergravity.

\begin{figure}[h]
\center
  \includegraphics[width=5.5in]{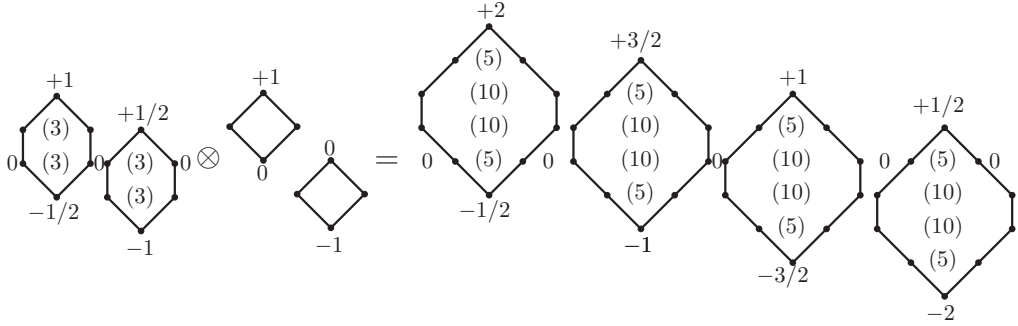}\\
  \caption{Diamond diagram for (supergravity)$_{\mathcal{N}_G=5}$
= (super Yang-Mills)$_{\widetilde{\mathcal{N}}=3}\otimes$
(super Yang-Mills)$_{\mathcal{N}=2}$. The four diamonds represent the four
superfields $\Phi^{\mathcal{N}_G=5}$,
 $\Theta^{\mathcal{N}_G=5}$, $\Gamma^{\mathcal{N}_G=5}$
and $\Psi^{\mathcal{N}_G=5}$. The hidden
indices are $(4)$ for $\Theta$, $(78)$ for $\Gamma$ and $(478)$ for
$\Psi$.}\label{klt523}
\end{figure}

The second case is obtained by tensoring $\widetilde{\mathcal{N}}=3$ with
$\mathcal{N}=2$, as shown in figure \ref{klt523}. There are four
diamonds, decreasing in helicity by half a unit. By recovering the hidden
indices from the diagrams, one sees immediately that we can write
\bea
\Phi^{\mathcal{N}_G=6}=\Phi^{\mathcal{N}_G=5}+
\eta_4\Theta^{\mathcal{N}_G=5}~,~~~
\Psi^{\mathcal{N}_G=6}=\Gamma^{\mathcal{N}_G=5}+\eta_4\Psi^{\mathcal{N}_G=5}~.~~~
\eea
So the four superfields of this theory are just a rewriting of the two
superfields $\Phi^{\mathcal{N}_G=6}$ and $\Psi^{\mathcal{N}_G=6}$ of minimal $\mathcal{N}_G=6$ supergravity.

\subsubsection{Diamond diagrams for the $\mathcal{N}_G=4$ theories}

\begin{figure}[h]
\center
  \includegraphics[width=4in]{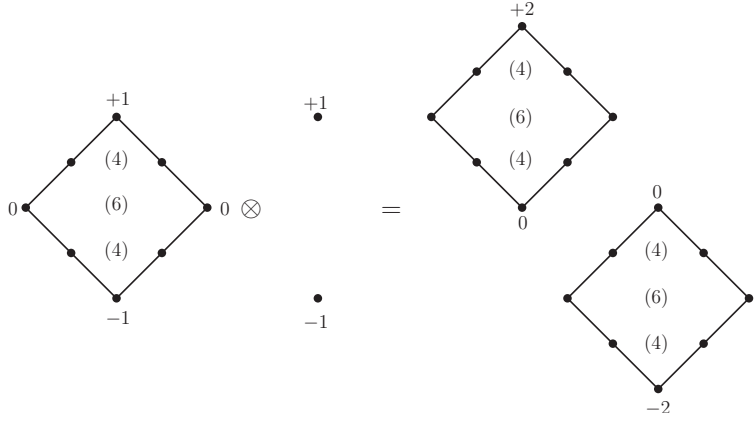}\\
  \caption{Diamond diagram for (supergravity)$_{\mathcal{N}_G=4}$
= (super Yang-Mills)$_{\widetilde{\mathcal{N}}=4}\otimes$
(Yang-Mills)$_{\mathcal{N}=0}$. The two diamonds represent
the superfields $\Phi^{\mathcal{N}_G=4}$ and $\Psi^{\mathcal{N}_G=4}$.
There are hidden indices $(5678)$ for the $\Psi$ field.
}\label{klt404}
\end{figure}

In this case we have three KLT constructions given by $(\widetilde{\mathcal{N}}=4)\otimes(\mathcal{N}=0)$,
$(\widetilde{\mathcal{N}}=3)\otimes(\mathcal{N}=1)$ and
$(\widetilde{\mathcal{N}}=2)\otimes(\mathcal{N}=2)$. The first
one is shown in figure \ref{klt404}. It contains only a
supergravity multiplet and its CPT conjugate, with a total of 1 graviton $h_{\pm}$, 4 gravitinos
$\psi_{\pm}$, 6 vectors $v_{\pm}$, 4 spin-1/2 fermions $\chi_{\pm}$ and two
scalars. This is minimal $\mathcal{N}_G=4$ supergravity. It was this
particular KLT-map that was recently used by Bern et al. to study finiteness
properties of $\mathcal{N}_G=4$ supergravity \cite{Bern5}.

\begin{figure}[h]
\center
  \includegraphics[width=5.5in]{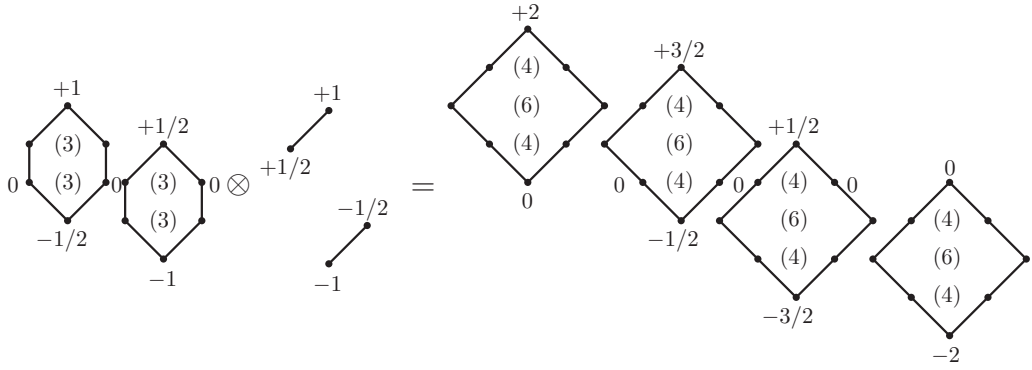}\\
  \caption{Diamond diagram for (supergravity)$_{\mathcal{N}_G=4}$
= (super Yang-Mills)$_{\widetilde{\mathcal{N}}=3}\otimes$
(super Yang-Mills)$_{\mathcal{N}=1}$. The four diamonds represent the
$\Phi^{\mathcal{N}_G=4}$, $\Theta^{\mathcal{N}_G=4}$,
$\Gamma^{\mathcal{N}_G=4}$ and $\Psi^{\mathcal{N}_G=4}$
superfields.
The hidden indices are $(4)$ for $\Theta$, $(678)$ for $\Gamma$ and
$(4678)$ for the $\Psi$ superfield. }\label{klt413}
\end{figure}

The second construction is illustrated in figure \ref{klt413}. As there is
just half a unit of helicity between both the
$\Phi^{\mathcal{N}_G=4}$ and $\Theta^{\mathcal{N}_G=4}$
superfields and the $\Gamma^{\mathcal{N}_G=4}$ and
$\Psi^{\mathcal{N}_G=4}$ superfields, we can readily recover the hidden indices,
and reassemble these four superfields into
\bea
\Phi^{\mathcal{N}_G=5}=\Phi^{\mathcal{N}_G=4}+
\eta_4\Theta^{\mathcal{N}_G=4}~,~~~
\Psi^{\mathcal{N}_G=5}=\Gamma^{\mathcal{N}_G=4}+\eta_4\Psi^{\mathcal{N}_G=4}~.~~~
\eea
Indeed, this theory is nothing but a rewriting of minimal
$\mathcal{N}_G=5$ supergravity.

\begin{figure}[h]
\center
  \includegraphics[width=5.5in]{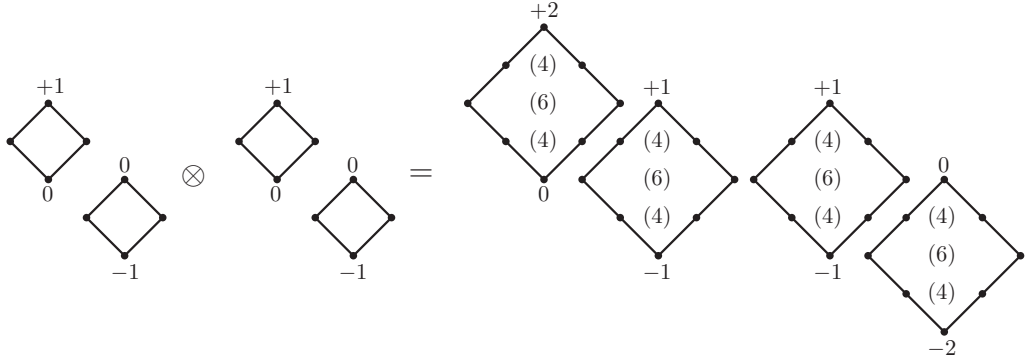}\\
  \caption{Diamond diagram for (supergravity)$_{\mathcal{N}_G=4}$
= (super Yang-Mills)$_{\widetilde{\mathcal{N}}=2}\otimes$
(super Yang-Mills)$_{\mathcal{N}=2}$. The four diamonds represent the
$\Phi^{\mathcal{N}_G=4}$, $\Theta^{\mathcal{N}_G=4}_{vector}$, $\Gamma^{\mathcal{N}_G=4}_{vector}$
and $\Psi^{\mathcal{N}_G=4}$ superfields. $\Theta$ and $\Gamma$ correspond to
CPT self-conjugate vector multiplets, but they have different sets
of $SU(4)_R$ indices. The hidden indices are $(34)$ for $\Theta$,
$(78)$ for $\Gamma$ and $(3478)$ for $\Psi$.}\label{klt422}
\end{figure}

The third construction is shown in figure \ref{klt422}. In contrast to the
result of the previous two cases, this theory contains two extra
diamonds besides the minimal supergravity multiplet. These two diamonds
of additional states cannot be reassembled
into new superfields with higher $\mathcal{N}_G$. Therefore, for the first
time in our systematic approach we here have a case where in addition to
a simple supergravity the KLT-map also provides us with a matter multiplet.
In detail, besides the field content of the
$(\widetilde{\mathcal{N}}=4)\otimes(\mathcal{N}=0)$ theory, the
additional matter fields combine to form a vector multiplet. It consists of
2 vector fields with helicity $\pm 1$, 8 fermion
fields of helicity $\pm 1/2$ and 12 scalars. The resulting theory
is $\mathcal{N}_G=4$ supergravity coupled to these matter fields.

%

\subsubsection{Diamond diagrams for the $\mathcal{N}_G=3$ theories}

\begin{figure}[h]
\center
  \includegraphics[width=4.5in]{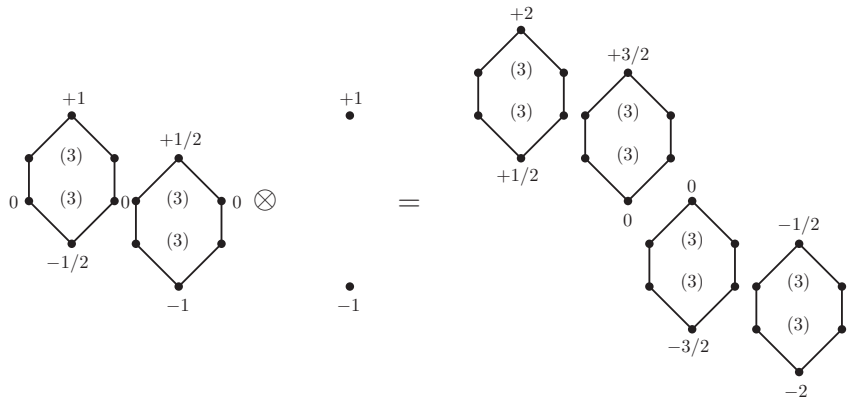}\\
  \caption{Diamond diagram for (supergravity)$_{\mathcal{N}_G=3}$
= (super Yang-Mills)$_{\widetilde{\mathcal{N}}=3}\otimes$
(Yang-Mills$_{\mathcal{N}=0}$. The four diamonds represent the four superfields
$\Phi^{\mathcal{N}_G=3}$, $\Theta^{\mathcal{N}_G=3}$, $\Gamma^{\mathcal{N}_G=3}$
and $\Psi^{\mathcal{N}=3}$. The hidden indices are
$(4)$ for $\Theta$, $(5678)$ for $\Gamma$ and $(45678)$ for $\Psi$.
}\label{klt303}
\end{figure}

There are two cases to consider. The first comes
from the product $(\widetilde{\mathcal{N}}=3)\otimes
(\mathcal{N}=0)$, which is illustrated in figure \ref{klt303}.
Superficially this appears to be a supergravity multiplet coupled to a
gravitino supermultiplet. However, we notice that the first two diamonds
have a helicity difference of half a unit, likewise for the last two diamonds.
Thus we can reconstruct each of them into a bigger superfield by
recovering the hidden index, \textit{i.e.}
\bea
\Phi^{\mathcal{N}_G=4}=\Phi^{\mathcal{N}_G=3}+\eta_4\Theta^{\mathcal{N}_G=3}~,~~~
\Psi^{\mathcal{N}_G=4}=\Gamma^{\mathcal{N}_G=3}+\eta_4\Psi^{\mathcal{N}_G=3}~.~~~
\eea
The resulting theory is simply minimal
$\mathcal{N}_G=4$ supergravity encoded in a slightly different way. This is
as expected since $\mathcal{N}=3$ super
Yang-Mills can be identified with $\mathcal{N}=4$ super Yang-Mills
theory.

\begin{figure}[h]
\center
  \includegraphics[width=4.5in]{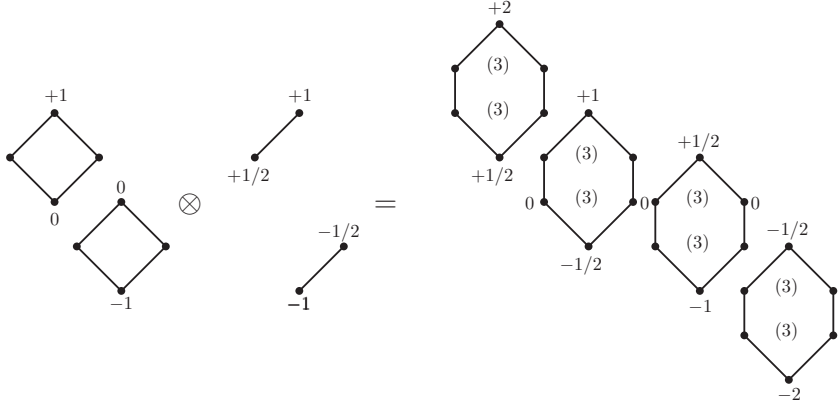}\\
  \caption{Diamond diagram for (supergravity)$_{\mathcal{N}_G=3}$
= (super Yang-Mills)$_{\widetilde{\mathcal{N}}=2}\otimes$
(super Yang-Mills)$_{\mathcal{N}=1}$. The four diamonds represent the
$\Phi^{\mathcal{N}_G=3}$, $\Theta^{\mathcal{N}_G=3}_{vector}$, $\Gamma^{\mathcal{N}_G=3}_{vector}$
and $\Psi^{\mathcal{N}_G=3}$ superfields. The hidden indices are
$(34)$ for $\Theta$, $(678)$ for $\Gamma$ and $(34678)$ for $\Psi$.
}\label{klt312}
\end{figure}

The second case is $(\widetilde{\mathcal{N}}=2)\otimes
(\mathcal{N}=1)$, see figure \ref{klt312}.
Besides the usual field content of minimal $\mathcal{N}_G=3$
supergravity, this theory has additional fields from the
matter supermultiplet, which includes 1 vector field $\pm 1$, 4
fermion fields $\pm 1/2$ and 6 scalars. This is a theory of minimal
$\mathcal{N}_G=3$ supergravity coupled to a vector mutliplet.

\subsubsection{Diamond diagrams for the $\mathcal{N}_G=2$ theories}

%
\begin{figure}[h]
\center
  \includegraphics[width=4.5in]{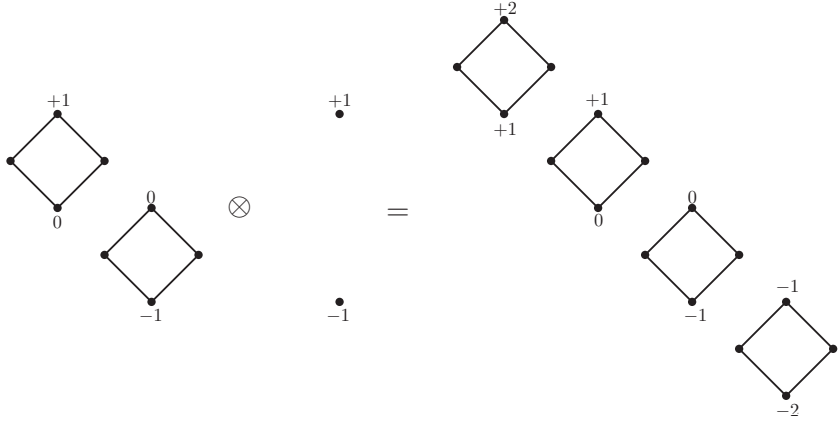}\\
  \caption{Diamond diagram for (supergravity)$_{\mathcal{N}_G=2}$
= (super Yang-Mills)$_{\widetilde{\mathcal{N}}=2}\otimes$
(Yang-Mills)$_{\mathcal{N}=0}$. The four diamonds represent $\Phi^{\mathcal{N}_G=2}$,
$\Theta^{\mathcal{N}_G=2}_{vector}$, $\Gamma^{\mathcal{N}_G=2}_{vector}$
and $\Psi^{\mathcal{N}_G=2}$. The hidden indices are $(34)$ for
$\Theta$, $(5678)$ for $\Gamma$ and $(345678)$ for $\Psi$.
}\label{klt202}
\end{figure}
There are again two constructions to consider. The first one comes from the product
$(\widetilde{\mathcal{N}}=2)\otimes (\mathcal{N}=0)$, as shown in
figure \ref{klt202}. This gives a theory containing a minimal $\mathcal{N}_G=2$
supergravity multiplet coupled to a vector multiplet. The
supergravity multiplet contains 1 graviton $h_{\pm}$, 2 gravitinos $\psi_{\pm}$
and 1 vector $\chi_{\pm}$, while the vector multiplet
contains 1 vector field of helicity $\pm 1$, 2 fermion fields of helicity $\pm 1/2$ and two
scalars.
\begin{figure}[h]
\center
  \includegraphics[width=4.5in]{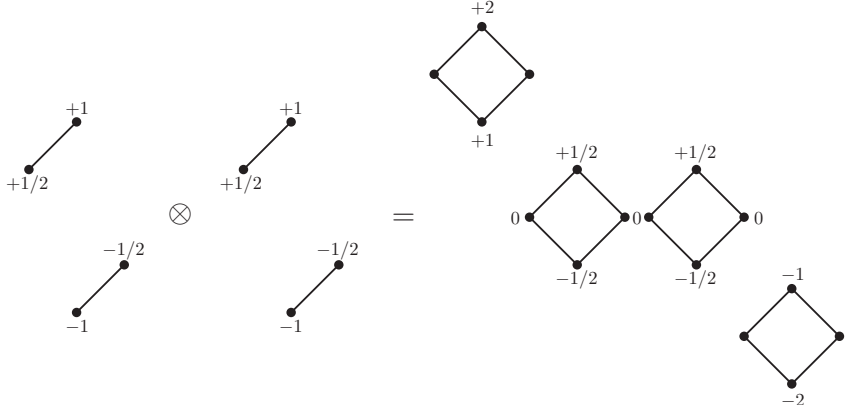}\\
  \caption{Diamond diagram for (supergravity)$_{\mathcal{N}_G=2}$
= (super Yang-Mills)$_{\widetilde{\mathcal{N}}=1}\otimes$
(super Yang-Mills)$_{\mathcal{N}=1}$. The four diamonds represent
the superfields
$\Phi^{\mathcal{N}_G=2}$, $\Theta^{\mathcal{N}_G=2}_{hyper}$,
$\Gamma^{\mathcal{N}_G=2}_{hyper}$ and $\Psi^{\mathcal{N}_G=2}$.
Hidden indices are $(234)$ for $\Theta$, $(678)$ for $\Gamma$ and
$(234678)$ for $\Psi$.}\label{klt211}
\end{figure}

The second construction is given by the product of
$(\widetilde{\mathcal{N}}=1)\otimes (\mathcal{N}=1)$, as illustrated in
figure \ref{klt211}. This provides also a theory of minimal $\mathcal{N}_G=2$ supergravity
coupled to matter, but now to a hypermultiplet
$\Theta^{\mathcal{N}_G=2}_{hyper}$, $\Gamma^{\mathcal{N}_G=2}_{hyper}$,
which is different from the vector multiplet of the first
construction. Besides the usual field content of the minimal
$\mathcal{N}_G=2$ supergravity multiplet, there are now also 2 fermion
fields of helicity $\pm 1/2$ and 4 scalars from the hypermultiplet.

\subsubsection{Diamond diagrams for the $\mathcal{N}_G=1$ theory}
%
\begin{figure}[h]
\center
  \includegraphics[width=3in]{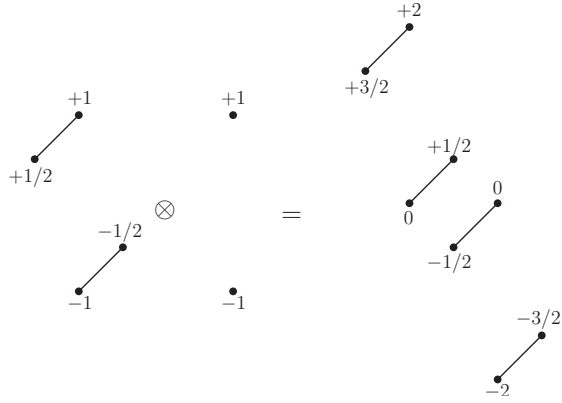}\\
  \caption{Diamond diagram for (supergravity)$_{\mathcal{N}_G=1}$
= (super Yang-Mills)$_{\widetilde{\mathcal{N}}=1}\otimes$
(Yang-Mills)$_{\mathcal{N}=0}$. The four diamonds represent
$\Phi^{\mathcal{N}_G=1}$, $\Theta^{\mathcal{N}_G=1}_{chiral}$, $\Gamma^{\mathcal{N}_G=1}_{chiral}$
and $\Psi^{\mathcal{N}_G=1}$. The hidden indices are $(234)$ for
$\Theta$, $(5678)$ for $\Gamma$ and $(2345678)$ for
$\Psi$.}\label{klt101}
\end{figure}
There is only one KLT-construction for this theory, namely
$(\widetilde{\mathcal{N}}=1)\otimes (\mathcal{N}=0)$, and it is shown
in figure \ref{klt101}. This is a theory of minimal $\mathcal{N}_G=1$
supergravity coupled to a hypermultiplet. There is 1 graviton
$h_{\pm}$ and 1 gravitino $\psi_{\pm}$ from the supergravity multiplet, and $1$
helicity-1/2 fermion field, as well as 2 scalars from the hypermultiplet. We can
work out all the states from tensor product of the (super) Yang-Mills theories, while
keeping track of both $R$-indices and hidden $R$-indices. We have
\bean &&\Phi^{\mathcal{N}_G=1}~:~~~(+1)\otimes (+1)=+2~~,~~(+{1\over
2}^{1})\otimes (+1)=+{3\over 2}^1~,~~\nonumber\\
&&\Theta^{\mathcal{N}_G=1}_{chiral}~:~~~(-{1\over 2}^{(234)})\otimes
(+1)=+{1\over
2}^{(234)}~~,~~(-1^{1(234)})\otimes(+1)=0^{1(234)}~,~~\nonumber\\
&&\Gamma^{\mathcal{N}_G=1}_{chiral}~:~~~(+1)\otimes(-1^{(5678)})=0^{(5678)}~~,~~(-{1\over
2}^{1})\otimes(-1^{(5678)})=-{1\over
2}^{1(5678)}~,~~\nonumber\\
&&\Psi^{\mathcal{N}_G=1}~:~~~(-{1\over 2}^{(234)})\otimes
(-1^{(5678)})=-{3\over
2}^{(2345678)}~,~~(-1^{1(234)})\otimes(-1)^{(5678)}=-2^{1(2345678)}~,~~
\eean
and all states are thus completely identified.

\subsubsection{Diamond diagrams for the $\mathcal{N}_G=0$ theory}

%
\begin{figure}[h]
\center
  \includegraphics[width=3in]{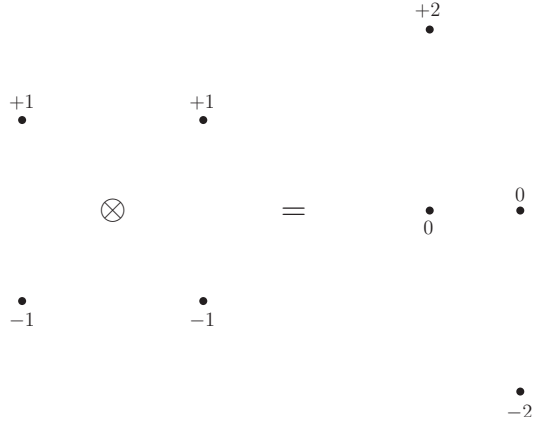}\\
  \caption{``Diamond" diagram for (gravity)$_{\mathcal{N}_G=0}$
= (Yang-Mills)$_{\widetilde{\mathcal{N}}=0}\otimes$
(Yang-Mills)$_{\mathcal{N}=0}$. This is just the classic case of gravity as the square of
two Yang-Mills theories. The origin of the two additional scalars is clear in the
language of hidden indices.}\label{klt000}
\end{figure}
For the $(\widetilde{\mathcal{N}}=0)\otimes (\mathcal{N}=0)$ product
the diamonds are reduced to simple dots, $i.e.$ single states. We can
still draw the ``diamond diagram" shown in figure \ref{klt000}.
This is Einstein gravity coupled to two scalars. The two
scalars come from tensor product of $(+1)\otimes
(-1^{(5678)})$, $(-1^{(1234)})\otimes (+1)$. Although the scalars do not have
explicit $R$-indices, they do carry different hidden $R$-indices.

\section{Symmetry groups of supergravity theories from super KLT-relations}
In this section, we will discuss the symmetry groups of the different
supergravity theories that can be constructed from KLT-products. Much of this
may be found in the supergravity literature, but it is quite instructive to see
the emergence of supergravity symmetries from the tensoring of two
super Yang-Mills theories.
\subsection{Maximal supersymmetric theories}

For supersymmetric theories with $\mathcal{N}$ supercharge generators $Q^A$,
$\widetilde{Q}_A$, $A=1,2,\ldots, \mathcal{N}$, it
  is possible to have an
$U(\mathcal{N})_R$ group that rotates $Q^A$ or $\widetilde{Q}_A$. It is customary
to decompose the $U(\mathcal{N})_R$ into
$U(\mathcal{N})_R=SU(\mathcal{N})_R\otimes U(1)_R$. The Grassmann
variables $\eta_A$ transform under $SU(\mathcal{N})_R$, and
the superamplitudes should be invariant under such a transformation.
This gives constraints on the $SU(\mathcal{N})_R$ indices of a given
superamplitude. For both supergravity and super Yang-Mills theories,
the $SU(\mathcal{N})_R$ group will surely be an invariant.

Let us now first briefly review the lack of an additional non-trivial $U(1)_R$ group
in maximally supersymmetric
theories. If there would be a non-trivial $U(1)_R$ group for the
$\mathcal{N}=4$ super Yang-Mills theory,
then at least some of the component fields should carry non-vanishing
$U(1)_R$ charges. It is  obvious that we cannot assign
non-trivial $U(1)_R$ charges to the gauge bosons (and such charges would
 be in contradiction with the existence of non-vanishing
pure-gluon MHV amplitudes). We could also try to assign
a $U(1)_R$ charge $\beta$ for the scalars, but recall that the superfield for
$\mathcal{N}=4$ super Yang-Mills theory is CPT self-conjugate. In our terminology,
all states are inside one single diamond, as shown in figure \ref{diamond}.
The scalars in this diamond satisfy a self-duality relation under complex conjugation.
As a consequnce, if we assign, say, $U(1)_R$ charge $\beta$ to the scalar $\phi^{12}$, then
the scalar $\phi^{34}=(\phi^{12})^\dagger$ will carry a $U(1)_R$ charge of
$-\beta$. Being states inside the same diamond, they should carry the same charge.
This gives $\beta=-\beta$, and forces all potential $U(1)_R$ charges
for the scalars to vanish. Finally,  could there be non-trivial $U(1)_R$ charges
for the fermions? The Yukawa couplings to the scalars forbid these, since the scalars
are neutral. From this argument one concludes that even if there is
potentially a $U(1)_R$ symmetry group in $\mathcal{N}=4$ super
Yang-Mills theory, all component fields have vanishing charges under this
symmetry and it therefore plays no role.
For a review of global symmetries in $\mathcal{N}=4$ super Yang-Mills theory, see \cite{Sohnius:1985qm}.

A similar argument carries through for the maximally supersymmetric supergravity theory.
Also here a crucial ingredient is that the
superfield of $\mathcal{N}_G=8$ supergravity is CPT self-conjugate,
so all states belong to one single diamond as shown in figure
\ref{klt844}. The $U(1)_R$ charge for the graviton must be zero.
The $U(1)_R$ charges for the scalars should also be zero, because scalars
and their complex-conjugate partners are all inside the same diamond; they
must hence carry the same $U(1)_R$ charges, which clearly forces all these charges to
vanish. For the remaining fields we need to look at their interactions. In the present
setting, this is actually most easily done by evaluating on-shell three-point amplitudes,
which directly correspond to interaction vertices in the supergravity Lagrangian.
In maximally supersymmetric supergravity, we find non-vanishing
three-point amplitudes such as
\begin{equation}
  \mathcal{M}(v^-,\chi^-,\chi^-) =\frac{\langle 12 \rangle^3 \langle 13 \rangle^3
    \langle 23 \rangle^2}{\langle 12 \rangle^2 \langle 23 \rangle^2
    \langle 31 \rangle^2}=\langle 12\rangle \langle 13\rangle~,~~~
\end{equation}
which represents two Weyl spinors coupled to a graviphoton. Indeed, this is
what in the Lagrangian corresponds to a
helicity-changing three-vertex through a gauge invariant Pauli term.

There is also
\begin{equation}
  \mathcal{M} (\phi,\chi^-,\psi^-) =\frac{\langle 12 \rangle \langle 13 \rangle^3
    \langle 23 \rangle^4}{\langle 12 \rangle^2 \langle 23 \rangle^2
    \langle 31 \rangle^2}=\langle 13\rangle \langle
  23\rangle^2/\langle 12 \rangle~,~~~
\end{equation}
which is the derivative coupling for a scalar field to a gravitino
and a spin-1/2 fermion. And another three-vertex corresponds to
\begin{equation}
  \mathcal{M} (\phi,v^-,v^-) =\frac{\langle 12 \rangle^2 \langle 13 \rangle^2
    \langle 23 \rangle^4}{\langle 12 \rangle^2 \langle 23 \rangle^2
    \langle 31 \rangle^2}=\langle   23\rangle^2~,~~~
\end{equation}
which is a scalar coupled to two graviphotons. Since all of these
amplitudes (and their corresponding vertices) are
non-vanishing, their total $U(1)_R$  charges must be zero. From
the vanishing charge of the scalar and three-vertex $\mathcal{M}
(\phi,v^-,v^-)$ we know that the $U(1)_R$ charge for graviphoton $v$
with helicity $-1$ should be zero. By further taking into account
the three-vertex corresponding to the amplitudes
$\mathcal{M}(v^-,\chi^-,\chi^-)$ we assure that the
$U(1)_R$ charge for graviphotino $\chi$ with helicity $-1/2$ should
be zero. Finally, from the three-vertex corresponding to
$\mathcal{M} (\phi,\chi^-,\psi^-)$ we
conclude that $U(1)_R$ charge for the gravitino $\psi$ of helicity
$-3/ 2$ should be zero. The same vanishing $U(1)_R$ charge must clearly
be assigned to their complex conjugate partners. From this simple argument,
we conclude that even if there is a potential $U(1)_R$ symmetry group
for maximally supersymmetric supergravity,
all the component fields will have zero $U(1)_R$ charge. So
such a group can play no role in this theory either.

The global linear symmetry groups of $\mathcal{N}_G=8$ supergravity and $\mathcal{N}=4$
super Yang-Mills theory are therefore $SU(8)$
and $SU(4)$, respectively.  For $\mathcal N_G=8$ supergravity, there are
also non-linear global symmetries, and the total global symmetry group
is $E_{7(7)}$ \cite{Cremmer:1978ds}. However, the non-linear symmetries do not generate
vanishing identities and they only manifest themselves in the soft
limit of scattering amplitudes \cite{ArkaniHamed:2008gz}.

\subsection{$4\leq \mathcal{N}_G<8$ minimal supergravity}

For the minimal $\mathcal{N}_G=4,5,6$ supergravity theories, there are only
supergravity multiplets, and therefore, in our notation, only the two
diamonds representing the $\Phi$ and $\Psi$ fields. Again there is a
$U(\mathcal{N}_G)_R$ group that
rotates $Q^A$ and $\widetilde{Q}_A$, where
$A=1,2,\ldots,\mathcal{N}_G$. The $SU(\mathcal{N}_G)_R$ group will for sure be an
invariant group for the amplitudes, but what about the
$U(1)_R$ group? Let us take the $\mathcal{N}_G=6$ minimal supergravity theory as
an example, and discuss the role of $U(1)_R$ symmetry in such minimal
supergravity theories.

The diamond diagram for $\mathcal{N}_G=6$ minimal supergravity is shown in
figure \ref{klt624}. We can read off the explicit
  $U(1)_R$ symmetry of $\mathcal
  N_G=6$ by truncating the $\mathcal N_G=8$ theory. 
We remind ourselves that the component
fields not only carry $SU(6)_R$ indices but also the hidden
$(78)$ indices. These hidden indices will appear in the component
fields of the $\Psi$ superfield, and they always come out neatly joined.
The generators of the symmetry group acting on the $R$-indices in the
$\mathcal{N}_G=6$ theory are $6\times 6$
matrices. They  are embedded in the Lie-algebra of $SU(8)$. Consider the following $8\times 8$
traceless Hermitian matrix
\bea \left(
       \begin{array}{cccccccc}
         \alpha & 0 & 0 & 0 & 0 & 0 & ~& ~ \\
         0 & \alpha & 0 & 0 & 0 & 0 & ~ & ~ \\
         0 & 0 & \alpha & 0 & 0 & 0 & ~ & ~\\
         0 & 0 & 0 & \alpha & 0 & 0 &  ~ & ~ \\
         0 & 0 & 0 & 0 & \alpha & 0 & ~ &  ~ \\
         0 & 0 & 0 & 0 & 0 & \alpha & ~  &  ~ \\
         ~ & ~ & ~ & ~ & ~ & ~ & B_{77} & B_{78} \\
         ~ & ~ & ~ & ~ & ~ & ~ & B_{87} & B_{88} \\
       \end{array}
     \right)~,~~~\eea
which is divided into two blocks. The first $6\times 6$ block acts
on the $R$-indices $1,2,\ldots, 6$ and the second $2 \times 2$ block
$B_{ij}$ acts on the hidden indices $(78)$. Since we are considering
the role of the $U(1)_R$ symmetry, we take
the first block to commute with all $SU(6)_R$
generators. It must therefore be a diagonal matrix proportional to the
identity, and we write it as $\alpha I_{6\times 6}$
in the
above matrix. This will
make each $R$-index $1,2,\ldots, 6$ correspond to a charge $\alpha$. What
is the effect of $B_{ij}$ acting on the hidden indices $(78)$?

For a more
general discussion, we will consider the effect of a $k\times k$
matrix $B_{x_i,x_j}$ acting on the indices
$(x_1,x_2,\ldots,x_k)$. Since Grassmann variables $\eta$ are totally
anti-commuting, this actually means a $k\times k$ matrix $B_{ij}$ acting
on $\eta_{x_1}\wedge\eta_{x_2}\wedge\cdots\wedge\eta_{x_k}$, where we
consider the $k$ Grassmann variables $\eta$ as spanning a superspace, with
each $\eta_{x_i}$ being a basis vector
$\eta_{x_i}=(0,\ldots,0,1,0,\ldots,0)^T$, and the $1$ is in the $i$-th
position. Then the matrix $B_{ij}$ acting on the basis vector
$\eta_{x_i}$ gives 
\bea \left(
       \begin{array}{ccccc}
         B_{11} & \cdots & \cdots & \cdots & B_{1k} \\
         \vdots & \ddots & ~ & ~ & \vdots \\
         \vdots & ~ & \ddots & ~ & \vdots \\
         \vdots & ~ & ~ & \ddots & \vdots \\
         B_{k1} & \cdots & \cdots & \cdots & B_{kk} \\
       \end{array}
     \right)_{k\times k}\left(
                                \begin{array}{c}
                                  0 \\
                                  \vdots \\
                                  1 \\
                                  \vdots \\
                                  0 \\
                                \end{array}
                              \right)=\left(
                                        \begin{array}{c}
                                          B_{1i} \\
                                          B_{2i} \\
                                          \vdots \\
                                          \vdots \\
                                          B_{ki} \\
                                        \end{array}
                                      \right)=\sum_{j=1}^{k}B_{ji}\eta_{x_j}~,~~~\eea
so
\bean
B(\eta_{x_1}\wedge\eta_{x_2}\wedge\cdots\wedge\eta_{x_k})&=&
\sum_{i=1}^{k}\eta_{x_1}\wedge\cdots\wedge(B\eta_{x_i})\wedge\cdots\wedge\eta_{x_k}\nonumber\\
&=&\sum_{i=1}^{k}\eta_{x_1}\wedge\cdots\wedge(\sum_{j=1}^{k}
B_{ji}\eta_{x_j})\wedge\cdots\wedge\eta_{x_k}\nonumber\\
&=&\sum_{i=1}^{k}\eta_{x_1}\wedge\cdots\wedge(B_{ii}\eta_{x_i})
\wedge\cdots\wedge\eta_{x_k}\nonumber\\
&=&(\sum_{i=1}^kB_{ii})(\eta_{x_1}\wedge\eta_{x_2}\wedge\cdots\wedge\eta_{x_k})~,~~~\eean
which means that only the trace part of the matrix $B_{ij}$ is important
when acting on the hidden indices. Note that this is true only when
the hidden indices appear together, so that they can be
expressed as
$(\eta_{x_1}\wedge\eta_{x_2}\wedge\cdots\wedge\eta_{x_k})$. If we
denote $\beta=Tr(B_{ij})$, we can then assign a charge of $\beta$
to the hidden indices $(x_1,x_2,\ldots,x_k)$.

Let us now return to the $\mathcal{N}_G=6$ minimal supergravity theory. For
each $R$-index $1,2,\ldots,6$ we assigned a charge $\alpha$,
and for the hidden indices $(78)$ we assigned a charge of $\beta$. Because
of the traceless condition on $SU(8)$ generators we have $\beta=-6\alpha$. In
this setup we can hence assign charges for all component fields through
their $SU(6)_R$ and/or hidden indices. For the $\Phi$-superfield diamond
we have \vspace{0.2in}

\begin{tabular}{|c|l|c|}
  \hline
  Helicity & KLT Products & Charge \\
  \hline
  $+2$ & $(+1)\otimes (+1)$ & $0$ \\
 $+{3\over 2}$ &$(+{1\over 2}^{a_1})\otimes (+1)~~,~~(+1)\otimes (+{1\over 2}^{b_1})$  & $\alpha$ \\
  $+1$ & $(0^{a_1a_2})\otimes (+1)~~,~~
(+{1\over 2}^{a_1})\otimes (+{1\over 2}^{b_1})~~,~~(+1)\otimes (0^{56}) $& $2\alpha$ \\
  $+{1\over 2}$ & $(-{1\over 2}^{a_1a_2a_3})\otimes (+1)~~,~~(0^{a_1a_2})\otimes
(+{1\over 2}^{b_1})~~,~~(+{1\over 2}^{a_1})\otimes (0^{56})$ & $3\alpha$ \\
  $0$ & $(-1^{1234})\otimes (+1)~~,~~(-{1\over 2}^{a_1a_2a_3})\otimes
(+{1\over 2}^{b_1})~~,~~(0^{a_1a_2})\otimes (0^{56})$ & $4\alpha$ \\
  $-{1\over 2}$ & $(-1^{1234})\otimes (+{1\over 2}^{b_1})~~,~~
(-{1\over 2}^{a_1a_2a_3})\otimes (0^{56})$ & $5\alpha$ \\
  $-1$ & $(-1^{1234})\otimes (0^{56})$ & $6\alpha$ \\
  \hline
\end{tabular}
\vspace{0.2in}

where $a_i=1,2,3,4$, and $b_i=5,6$. For the $\Psi$ superfield we
have \vspace{0.2in}

\begin{tabular}{|c|l|c|}
  \hline
  Helicity & KLT Products & Charge \\
  \hline
    $+1$ & $(+1)\otimes (0^{(78)})$ & $-6\alpha$ \\
    $+{1\over 2}$ & $(+1)\otimes (-{1\over 2}^{b_1(78)})~~,~~
(+{1\over 2}^{a_1})\otimes (0^{(78)})$ & $-5\alpha$ \\
    $0$ & $(+1)\otimes (-1^{56(78)})~~,~~(+{1\over 2}^{a_1})\otimes
(-{1\over 2}^{b_1(78)})~~,~~(0^{a_1a_2})\otimes (0^{(78)})$ & $-4\alpha$ \\
    $-{1\over 2}$ & $(+{1\over 2}^{a_1})\otimes (-1^{56(78)})~~,~~
(0^{a_1a_2})\otimes (-{1\over 2}^{b_1(78)})~~,~~(-{1\over 2}^{a_1a_2a_3})\otimes (0^{(78)})$ & $-3\alpha$ \\
    $-1$ & $(0^{a_1a_2})\otimes (-1^{56(78)})~~,~~
(-{1\over 2}^{a_1a_2a_3})\otimes (-{1\over 2}^{b_1(78)})~~,~~(-1^{1234})\otimes (0^{(78)}) $& $-2\alpha$ \\
    $-{3\over 2}$ &$(-{1\over 2}^{a_1a_2a_3})\otimes (-1^{56(78)})~~,~~
(-1^{1234})\otimes (-{1\over 2}^{b_1(78)})$  & $-\alpha$ \\
    $-2$ & $(-1^{1234})\otimes (-1^{56(78)})$ & $0$ \\
  \hline
\end{tabular}
\vspace{0.2in}

Since the $\Psi$ superfield is the CPT-conjugate of the $\Phi$ superfield,
the charges of its component fields are opposite to the charges of
their CPT-conjugated component partners in the $\Phi$ superfield. This is
exactly as one can read off from the above two tables in our $\mathcal{N}_G=6$ example.
The charges of component fields within each
diamond are different, which just means that
this symmetry does not commute with supercharges. The freedom of choice of $\alpha$
corresponds to an additional $U(1)_R$ invariance group for
superamplitudes. Combined with the $SU(6)_R$ invariant group, we infer
that superamplitudes of $\mathcal{N}_G=6$ minimal supergravity must be
invariant under the larger $U(6)_R=SU(6)_R\otimes U(1)_R$
group.

Similarly, for minimal supergravity theories of $4\leq \mathcal{N}_G<8$,
there is always a freedom of assigning an abelian charge $\alpha$ to the component
fields of these theories, and the superamplitudes are therefore invariant
under the larger $U(\mathcal{N}_G)_R=SU(\mathcal{N}_G)_R\otimes U(1)_R$ group.

\subsection{$0\leq \mathcal{N}_G\leq 4$ minimal gravity coupled to matter multiplets}

For $0\leq \mathcal{N}_G\leq 4$ supergravity theories, there will, besides the
usual supergravity multiplets, also emerge matter supermultiplets
from the KLT product. In detail, we have $\Phi$-$\Psi$ superfields for the
supergravity multiplets, and $\Theta$-$\Gamma$ superfields for the
matter multiplets. The $U(\mathcal{N}_G)_R$ group that rotates
$Q^A$ and $\widetilde{Q}_A$ is still there, and the full $SU(\mathcal{N}_G)_R$
group is an invariant group for the superamplitudes. In order to
study other possible invariant groups for these kind of theories, let
us consider
$(\mathcal{N}_G=4)=(\widetilde{\mathcal{N}}=2)\otimes(\mathcal{N}=2)$
supergravity from the KLT-construction. The associated diamond diagrams are
shown in figure \ref{klt422}, and there are hidden indices $(34)$ from
the $\widetilde{\mathcal{N}}=2$ super Yang-Mills theory and $(78)$ from the other.
The $SU(4)_R$ indices for this $\mathcal{N}_G=4$ supergravity theory are then
$1,2,5,6$. Consider now the following $8 \times 8$  matrix 
in the Lie-algebra of $SU(8)$,
\bea \left(
       \begin{array}{cccccccc}
         \alpha & 0 & ~ & ~ & 0 & 0 & ~& ~ \\
         0 & \alpha & ~ & ~ & 0 & 0 & ~ & ~ \\
         ~ & ~ & B_{33} & B_{34} & ~ & ~ & ~ & ~\\
         ~ & ~ & B_{43} & B_{44} & ~ & ~ &  ~ & ~ \\
         0 & 0 & ~ & ~ & \alpha & 0 & ~ &  ~ \\
         0 & 0 & ~ & ~ & 0 & \alpha & ~  &  ~ \\
         ~ & ~ & ~ & ~ & ~ & ~ & C_{77} & C_{78} \\
         ~ & ~ & ~ & ~ & ~ & ~ & C_{87} & C_{88} \\
       \end{array}
     \right)~.~~~\eea
Since we are again looking for the effects of $U(1)$ subgroups,
we take the $4\times 4$ matrix of $\mathcal{N}_G=4$ supergravity, \textit{i.e.} the
matrix constructed from row $1,2,5,6$ and column $1,2,5,6$ of the above
$8\times 8$ matrix, to commute with all generators of
$SU(4)_R$. It will thus be a diagonal matrix proportional to the
identity $\alpha I_{4\times 4}$, as already written
above. $B_{ij}$ is a $2\times 2$ matrix acting on the hidden
indices $(34)$ while $C_{ij}$ is another $2\times 2$ matrix acting
on the hidden indices $(78)$. From the previous discussion we
know that only the trace part of $B_{ij}$ and $C_{ij}$
acts on the hidden indices. The indices $(34)$ and
$(78)$ are not paired, so we should treat the two matrices
separately. If we denote $\beta=Tr(B_{ij})$ and $\gamma=Tr(C_{ij})$,
then for each $SU(4)_R$ $R$-index $1,2,5,6$ we can assign a charge
$\alpha$, while for $(34)$ we assign a charge
$\beta$ and for $(78)$ a charge
$\gamma$. The condition of tracelessness of $SU(8)$ generators
implies the constraint $\gamma=-\beta-4\alpha$.

With this, we can now assign corresponding $U(1)$ charges for all component
fields in the $\Phi,\Theta,\Gamma$ and $\Psi$ superfields. More
explicitly, for the $\Phi$ superfield we have \vspace{0.2in}

\begin{tabular}{|c|l|c|}
  \hline
  Helicity & KLT Product & Charge \\
  \hline
  $+2$ & $(+1)\otimes (+1)$ & $0$ \\
  $+{3\over 2}$ & $(+{1\over 2}^{a_1})\otimes (+1)~~,~~
(+1)\otimes (+{1\over 2}^{b_1})$ & $\alpha$ \\
  $+1$ & $(0^{12})\otimes (+1)~~,~~(+{1\over 2}^{a_1})\otimes
(+{1\over 2}^{b_1})~~,~~(+1)\otimes (0^{56})$ & $2\alpha$ \\
  $+{1\over 2}$ & $(0^{12})\otimes (+{1\over 2}^{b_1})~~,~~
(+{1\over 2}^{a_1})\otimes (0^{56})$ & $3\alpha$ \\
  $0$ & $(0^{12})\otimes (0^{56})$ & $4\alpha$ \\
  \hline
\end{tabular}

\vspace{0.2in}

where $a_i=1,2$ and $b_i=5,6$. For the $\Theta$ superfield we have
\vspace{0.2in}

\begin{tabular}{|c|l|c|}
  \hline
  Helicity & KLT Product & Charge \\
  \hline
  $+1$ & $(0^{(34)})\otimes (+1)$ & $0+\beta$ \\
  $+{1\over 2}$ & $(0^{(34)})\otimes (+{1\over 2}^{b_1})~~,~~
(-{1\over 2}^{a_1(34)})\otimes (+1)$ & $\alpha+\beta$ \\
  $0$ & $(0^{(34)})\otimes (0^{56})~~,~~(-{1\over 2}^{a_1(34)})\otimes
(+{1\over 2}^{b_1})~~,~~(-1^{12(34)})\otimes (+1)$ & $2\alpha+\beta$ \\
  $-{1\over 2}$ & $(-{1\over 2}^{a_1(34)})\otimes (0^{56})~~,~~
(-1^{12(34)})\otimes (+{1\over 2}^{b_1})$ & $3\alpha+\beta$ \\
  $-1$ & $(-1^{12(34)})\otimes (0^{56})$ & $4\alpha+\beta$ \\
  \hline
\end{tabular}

\vspace{0.2in}

For the $\Gamma$-superfield \vspace{0.2in}

\begin{tabular}{|c|l|c|}
  \hline
  Helicity & KLT Product & Charge \\
  \hline
  $+1$ & $(+1)\otimes (0^{(78)})$ & $-4\alpha-\beta$ \\
  $+{1\over 2}$ & $(+1)\otimes (-{1\over 2}^{b_1(78)})~~,~~
(+{1\over 2}^{a_1})\otimes (0^{(78)})$ & $-3\alpha-\beta$ \\
  $0$ & $(0^{12})\otimes (0^{(78)})~~,~~(+{1\over 2}^{a_1})\otimes
(-{1\over 2}^{b_1(78)})~~,~~(+1)\otimes (-1^{56(78)})$ & $-2\alpha-\beta$ \\
  $-{1\over 2}$ & $(0^{12})\otimes (-{1\over 2}^{b_1(78)})~~,~~
(+{1\over 2}^{a_1})\otimes (-1^{56(78)})$ & $-\alpha-\beta$ \\
  $-1$ & $(0^{12})\otimes (-1^{56(78)})$ & $0-\beta$ \\
  \hline
\end{tabular}

\vspace{0.2in}

and finally for the $\Psi$-superfield \vspace{0.2in}

\begin{tabular}{|c|l|c|}
  \hline
  Helicity & KLT Product & Charge \\
  \hline
  $0$ & $(0^{(34)})\otimes (0^{(78)})$ & $-4\alpha$ \\
  $-{1\over 2}$ & $(0^{(34)})\otimes (-{1\over 2}^{b_1(78)})~~,~~
(-{1\over 2}^{a_1(34)})\otimes (0^{(78)})$ & $-3\alpha$ \\
  $-1$ & $(0^{(34)})\otimes (-1^{56(78)})~~,~~(-{1\over 2}^{a_1(34)})\otimes
(-{1\over 2}^{b_1(78)})~~,~~(-1^{12(34)})\otimes (0^{(78)})$ & $-2\alpha$ \\
  $-{3\over 2}$ & $(-{1\over 2}^{a_1(34)})\otimes (-1^{56(78)})~~,~~
(-1^{12(34)})\otimes (-{1\over 2}^{b_1(78)})$ & $-\alpha$ \\
  $-2$ & $(-1^{12(34)})\otimes (-1^{56(78)})$ & $0$ \\
  \hline
\end{tabular}

\vspace{0.2in}

{}From the charges of the component fields in these four superfields, we
see that (1) The charge of the graviton $h$ is indeed zero as it had to be,
(2) The $\Psi$ and $\Gamma$ superfields are CPT conjugate of the $\Phi$
and $\Theta$ superfields, respectively, and thus the charge of a given
component field is
opposite to its CPT-conjugate partner in the corresponding superfield,
(3) If we only consider the $\alpha$ charge and  set
  $\beta=0$, we
see that component fields in each diamond have different charges corresponding
to the usual $U(1)_R$ charge, (4) If we only consider the $\beta$
charge and set $\alpha=0$, then there is an extra charge for the
matter supermultiplet, and every component field inside the same
diamond has the same charge $\beta$ (or $-\beta$ in the $\Gamma$
diamond). This indicates that there is an extra $U(1)$ invariant
group for the matter multiplet that commutes with all supercharge
generators $Q^A,\widetilde{Q}_A$. This $U(1)$ group is different
from the $U(1)_R$ group that comes from $U(\mathcal{N}_G)_R$ rotations.

{}From the discussion above we learn that there is an $U(1)_R$ invariant
group from the freedom of assigning a charge $\alpha$ to the component
fields of all supermultiplets, and another $U(1)$ invariant group
from the freedom of assigning a charge $\beta$ to component fields in the
matter multiplets. We conclude that the superamplitudes of the $\mathcal{N}_G=4$
supergravity theory considered here is invariant under the group
$SU(4)_R\otimes U(1)_R\otimes U(1)$. Similarly, for $0\leq
\mathcal{N}_G<4$ supergravity theories that describe minimal
supergravity multiplets coupled to matter supermultiplets, the
superamplitudes are invariant under the group $U(\mathcal{N}_G)_R\otimes
U(1)=SU(\mathcal{N}_G)_R\otimes U(1)_R\otimes U(1)$.

\subsubsection{Examples of $SU(\mathcal{N}_G)_R\otimes U(1)_R\otimes U(1)$ symmetry}

In order to illustrate the impact of $SU(\mathcal{N}_G)_R$ and the additional
$U(1)_R\otimes U(1)$ group on gravity amplitudes, let us
start by considering some
amplitudes of gravitons coupled to two scalars in the $\mathcal{N}_G=4$
supergravity theory constructed from two $\mathcal{N}=2$ super Yang-Mills
amplitudes, see figure \ref{klt422}. The hidden
indices are $(34)$ and $(78)$, and $SU(4)_R$ $R$-indices are 1, 2, 5, 6. If
we assign a charge $\alpha$ to each of the four $SU(4)_R$ $R$-indices, and
$\beta$ for the hidden indices $(34)$, then the hidden indices $(78)$ will
carry a charge of $-4\alpha-\beta$ because of the traceless condition
of the generators of $SU(8)_R$. Let us take the following two scalars from the 
$\Theta^{\mathcal{N}_G=4}_{vector}$ diamond
\begin{align}
\phi_1 = (0^{(34)})\otimes (0^{56})\,, \qquad \phi_2 = (-1^{12(34)})\otimes (+1)\,,
\label{n2-scalar-1}
\end{align}
which carry
the $U(1)_R$ charge $2\alpha$ and $U(1)$ charge $\beta$. From the
$\Gamma_{vector}^{\mathcal{N}_G=4}$ superfield we pick
\begin{align}
\phi_3 = (0^{12})\otimes (0^{(78)})\,, \qquad \phi_4 = (+1)\otimes (-1^{56(78)})\,,
\label{n2-scalar-2}
\end{align}
with $U(1)_R$ charge $-2\alpha$ and $U(1)$ charge $-\beta$.
Consider now the following two-scalar MHV amplitude
\begin{equation}
M_n^{\mathcal{N}_G=2}(\phi_i,\phi_j,h^-,h^+,\ldots,h^+)~,~~~
\end{equation}
where the two scalars $\phi_i,\phi_j$, with $i,j=1,\ldots,4$, can be any one of the above four kinds of scalars in
(\ref{n2-scalar-1}) and (\ref{n2-scalar-2}). This amplitude
can be readily calculated from the KLT-relations, and the only two
non-vanishing amplitudes are
\begin{eqnarray}
M_n^{\mathcal{N}_G=4}(\phi_1,\phi_3,h^-,h^+,\ldots,h^+)~,~~~
M_n^{\mathcal{N}_G=4}(\phi_2,\phi_4,h^-,h^+,\ldots,h^+)~,~~~
\end{eqnarray}
which neither violates $SU(4)_R$ symmetry or the conservation of $U(1)_R\otimes
U(1)$ charge.
The first is constructed from products of two $\mathcal{N}=2$
super Yang-Mills MHV amplitudes with a scalar and its CPT-conjugated partner and all other legs being
gluons. The
second arises from two pure-gluon MHV amplitudes.
All other of these two-scalar
amplitudes vanishes, as can be easily inferred from the gauge theory
part. The super Yang-Mills amplitudes either contain only one scalar or two
of the same scalars.

These vanishing two-scalar supergravity amplitudes can be explained
from the violation of $SU(4)_R$ and/or $U(1)_R\otimes U(1)$ symmetry. For the following
amplitudes 
\begin{eqnarray}
&& M_n^{\mathcal{N}_G=4}(\phi_1,\phi_1,h^-,h^+,\ldots,h^+)~,~~~
M_n^{\mathcal{N}_G=4}(\phi_2,\phi_2,h^-,h^+,\ldots,h^+)~,~~~\nonumber \\
&& M_n^{\mathcal{N}_G=4}(\phi_3,\phi_3,h^-,h^+,\ldots,h^+)~,~~~
M_n^{\mathcal{N}_G=4}(\phi_4,\phi_4,h^-,h^+,\ldots,h^+)~,~~~
\end{eqnarray}
$SU(4)_R$ is violated as well as $U(1)_R\otimes U(1)$. For the amplitudes
\begin{align}
M_n^{\mathcal{N}_G=4}(\phi_1,\phi_2,h^-,h^+,\ldots,h^+)~,~~~
M_n^{\mathcal{N}_G=4}(\phi_3,\phi_4,h^-,h^+,\ldots,h^+)~,~~~
\end{align}
the $SU(4)_R$ is not violated, but the total $U(1)_R\otimes U(1)$
charge is non-zero, and thus they vanish.
For the amplitudes
\begin{align} 
M_n^{\mathcal{N}_G=4}(\phi_1,\phi_4,h^-,h^+,\ldots,h^+)~,~~~
M_n^{\mathcal{N}_G=4}(\phi_2,\phi_3,h^-,h^+,\ldots,h^+)~,~~
\end{align}
the $U(1)_R\otimes U(1)$ charge is zero, but $SU(4)_R$ is
violated.

The $U(1)_R$ and $U(1)$ groups can also be violated individually.
To illustrate this, let us have a look at some two-graviphoton coupled to graviton amplitudes.
In the $\Theta_{vector}^{\mathcal{N}_G=4}$ superfield, we have the graviphoton $v_1^+$ with charge $(0+\beta)$ and
$v_2^-$ with charge $(4\alpha+\beta)$, while in the $\Gamma_{vector}^{\mathcal{N}_G=4}$
superfield we have $v_2^+$ with charge $(-4\alpha-\beta)$ and
$v_1^-$ with charge $(0-\beta)$.

With this we can have non-vanishing amplitudes like
\begin{align}
M_n^{\mathcal{N}_G=4}(v_1^{+},v_1^{-},h^-,h^+,\ldots,h^+)~,~~~ M_n^{\mathcal{N}_G=4}(v_2^{+},v_2^{-},h^-,h^+,\ldots,h^+)\,,
\end{align}
which satisfy both the $SU(4)_R$ symmetry and conserves both the $U(1)_R$ and the $U(1)$ charge.
However, it is easy to see that, for example, the amplitude
\bea
M_n^{\mathcal{N}_G=4}(v_1^{-},v_1^{-},h^-,h^+,\ldots,h^+) ~,~~~
\eea
does not violate the $SU(4)_R$ symmetry and conserve the $U(1)_R$ charge,
but not the $U(1)$ charge. Thus the violation of $U(1)$
ensures the vanishing of this amplitude. Similarly, the amplitude
\bea
M_n^{\mathcal{N}_G=4}(v_1^{-},v_2^{-},h^-,h^+,\ldots,h^+)~,~~~
\eea
does not violate the $SU(4)_R$ symmetry, have zero $U(1)$ charge but the $U(1)_R$ charge is not conserved, thus the violation
of $U(1)_R$ implies the vanishing of this amplitude.

\subsection{Summary of the symmetry groups from the KLT-construction}

\begin{table}
\center

\begin{tabular}{|c|c|c|c|c|c|c|c|}
  \hline
  \multirow{2}{*}{$\mathcal{N}_G$} & \multirow{2}{*}{$\widetilde{\mathcal{N}}\otimes
\mathcal{N}$} & \multicolumn{5}{|c|}{Number of states for component
fields} &  Linear symmetry group\\
  \cline{3-7}
    & & 2 & 3/2 & 1 & 1/2 & 0 & from KLT product \\
  \hline
  8 & $4\otimes 4$ & 1 & 8 & 28 & 56 & 70 & $SU(8)_R$ \\
  \hline
  7 & $4\otimes 3$ & 1 & 7+1 & 21+7 & 35+21 & 35+35 & $SU(8)_R$ \\
  \hline
  6 & $3\otimes 3$ & 1 & 6+1+1 & 15+6+6+1 & 20+15+15+6 & 15+20+20+15 & $SU(8)_R$ \\
  \hline
  6 & $4\otimes 2$ & 1 & 6 & 15+1 & 20+6 & 15+15 & $U(6)_R$  \\
  \hline
   5 & $3\otimes 2$ & 1 & 5+1 & 10+5+1 & 10+10+5+1 & 5+10+10+5 & $U(6)_R$  \\
  \hline
   5 & $4\otimes 1$ & 1 & 5 & 10 & 10+1 & 5+5 & $U(5)_R$  \\
  \hline
   4 & $3\otimes 1$ & 1 & 4+1 & 6+4 & 4+6+1 & 1+4+4+1 & $U(5)_R$  \\
  \hline
   4 & $4\otimes 0$ & 1 & 4 & 6 & 4 & 1+1 & $U(4)_R$ \\
  \hline
  3 & $3\otimes 0$ & 1 & 3+1 & 3+3 & 1+3 & 1+1 & $U(4)_R$ \\
  \hline
   4 & $2\otimes 2$ & 1 & 4 & 6+1+1 & 4+4+4 & 1+6+6+1 & $U(4)_R\otimes U(1)$ \\
  \hline
  3 &  $2\otimes 1$ & 1 & 3 & 3+1 & 1+3+1 & 3+3 & $U(3)_R\otimes U(1)$ \\
  \hline
  2 & $2\otimes 0$ & 1 & 2 & 1+1 & 2 & 1+1 & $U(2)_R\otimes U(1)$ \\
  \hline
  2 & $1\otimes 1$ & 1 & 2 & 1 & 1+1 & 2+2 & $U(2)_R\otimes U(1)$ \\
  \hline
  1 & $1\otimes 0$ & 1 & 1 & 0 & 1 & 1+1 & $U(1)_R\otimes U(1)$ \\
  \hline
  0 & $0\otimes 0$ & 1 & 0 & 0 & 0 & 1+1 & $U(1)$ \\
  \hline

\end{tabular}

\caption{Field content of the supergravity theories that can be constructed from
super KLT-relations, and their invariant groups as inferred from our diamond
diagrams. The total number of states for specific component fields is
obtained by adding states in the different diamonds of the given theory.
The linear global symmetry groups for minimal $4\leq \mathcal{N}_G \leq 8$ supergravities are also listed in \cite{Cremmer:1979up}}
\label{table-group}
\end{table}

As shown in the examples discussed above, we see that the invariant
symmetry groups for supergravity theories constructed out of KLT-products
are directly linked to the type of diamonds for such a theory. In
general there is $U(\mathcal{N}_G)_R \sim SU(\mathcal{N}_G)_R$ and $U(1)_R$ symmetry.
For theories of maximal supersymmetry, or theories with exactly the same field
content as maximal supersymmetry, all component fields are forced to have vanishing
$U(1)_R$ charge, and thus $U(1)_R$ plays no role in these theories. For
theories with only supergravity multiplets which are not CPT self-conjugate,
component fields
in the same diamond will have different $U(1)_R$ charges, and
opposite $U(1)_R$ charges for the complex-conjugate partners in the
CPT-conjugate diamond. For theories that contains
matter supermultiplets, an extra $U(1)$ group appears naturally.
All component fields in the same
matter multiplet diamond have the same additional $U(1)$
charges, opposite to those  of the component fields in the CPT-conjugate
diamond. We list all possible supergravity theories constructed
from KLT-products of super Yang-Mills theories in table \ref{table-group},
along with the invariant symmetry group that can be inferred from the KLT-product.

Superamplitudes of the different supergravity theories are invariant
under their symmetry groups,  and both $SU(\mathcal{N}_G)_R$ and
$U(1)_R\otimes U(1)$ symmetries induce vanishing identities.

\section{Conclusions}

In this paper we have identified all possible four-dimensional KLT-maps between
gauge theories with decreasing supersymmetry to associated gravity theories
with correspondingly decreasing degree of supersymmetry. For the case of
$\mathcal N=0$, we recover the well-known map between pure Yang-Mills theory and
Einstein gravity. That map is actually slightly larger in that
also two scalars couple on the gravity side. These two scalars are the only remnants
of string theory in that simplest case: the axion and the dilaton. The existence of
this additional set of scalars and their corresponding conserved $U(1)$ quantum
number is the closest  we get to a conserved $R$-charge in that case.
Conservation of this charge is what ensures the existence of ``vanishing identities''
among the pure Yang-Mills amplitudes when helicities do not match in the
product of amplitudes.

For theories  with   maximal supersymmetry to
theories without supersymmetry,
we have derived the full catalog of equivalences between
gravity and gauge theories, and explored the linear symmetries that can be inferred
from the KLT-map. The pattern is  very interesting, and in particular for
supergravity theories with  $\mathcal N_G \leq 4$, there are additional
matter multiplets  (the two scalars coupled to gravity
in the ${\mathcal N}_G=0$ case can be viewed as an example of this phenomenon).
We have established the precise maps, with the aid of diamond
diagrams that graphically illustrates the combination of states on the gauge
theory side into supergravity states on the gravity side. Again, the complete
set of  linear global  symmetries  builds up the full set of vanishing identities.

Although the supergravity theories that follow from the gauge theory map
may contain additional matter multiplets, one can readily project out those by
fixing appropriate $R$-symmetry indices on the supergravity side.
In this sense
one can for, instance, construct minimal  ${\mathcal N}_G=4$ supergravity from both
$(\widetilde{{\mathcal N}}=4)\otimes ({\mathcal N}=0)$ 
directly {\em or} from $(\widetilde{{\mathcal N}}=2)\otimes ({\mathcal N}=2)$ together with the
projection discussed after eq.~\eqref{KLT_Cat_III}.

We have here restricted ourselves to KLT-maps between
pure Yang-Mills theories with varying
degrees of supersymmetry. It could be interesting to explore corresponding maps
based on Yang-Mills theories with matter fields as well.

\section*{Acknowledgment}
Numerous discussions with Emil Bjerrum-Bohr, Donal O'Connell, Henry Tye and Yi Yin
are gratefully acknowledged.

\appendix

\section{Explicit expressions for superfields of $\mathcal{N}_G<8$ supergravity\label{app1}}

In this appendix we present the explicit expressions for the superfields
we have represented by diamond diagrams throughout our paper.
The superfields of supergravity multiplets will be
denoted by $\Phi^{\mathcal{N}_G}$ and their CPT-conjugates by
$\Psi^{\mathcal{N}_G}$. The needed matter multiplets will be denoted
by $\Theta^{\mathcal{N}_G}$ and their CPT-conjugates by
$\Gamma^{\mathcal{N}_G}$.

For the $\mathcal{N}_G=7$ theory, we have the $(+2,+{3\over
2}^{7},+1^{21},+{1\over 2}^{35},0^{35},-{1\over
2}^{21},-1^{7},-{3\over 2})$ supergravity multiplet diamond
\begin{align} 
\Phi^{\mathcal{N}_G=7} = \Phi^{\mathcal{N}_G=8}|_{\eta_8
\rightarrow 0}
=&{} h_+ + \sum_{i=1}^7 \eta_i\psi_+^i +\sum_{i<j=1}^7
\eta_i\eta_j v_+^{ij} +\sum_{i<j<k=1}^7 \eta_i\eta_j\eta_k \chi_+^{ijk} \nonumber \\
& + \sum_{i<j<k<l=1}^7  \eta_i\eta_j\eta_k\eta_l \phi^{ijkl}
+\sum_{i<j<k<l<m=1}^7 \eta_i\eta_j\eta_k\eta_l\eta_m \chi_-^{ijklm} \nonumber \\
& + \sum_{i<j<k<l<m<p=1}^7\eta_i\eta_j\eta_k\eta_l\eta_m\eta_p v_-^{ijklmp}
+ \eta_1\eta_2\eta_3\eta_4\eta_5\eta_6\eta_7 \psi_-^{1234567}~,~~~ \nonumber \\
\end{align}
where the superscripts denote the the degeneracies of states. \\
We also have the
$(+{3\over 2},+1^{7},+{1\over 2}^{21},0^{35}, -{1\over
2}^{35},-1^{21},-{3\over 2}^{7},-2)$ supergravity multiplet diamond
\begin{align} 
\Psi^{\mathcal{N}_G=7} = \int  d\eta_8 \Phi^{\mathcal{N}_G=8}
=&{}
\psi_+^{(8)} - \sum_{i=1}^7 \eta_i v_+^{i(8)} +\sum_{i<j=1}^7
\eta_i\eta_j \chi_+^{ij(8)} -\sum_{i<j<k=1}^7 \eta_i\eta_j\eta_k
\phi^{ijk(8)} \nonumber\\
&+\sum_{i<j<k<l=1}^7  \eta_i\eta_j\eta_k\eta_l \chi_-^{ijkl(8)}
-\sum_{i<j<k<l<m=1}^7 \eta_i\eta_j\eta_k\eta_l\eta_m v_-^{ijklm(8)}
\nonumber\\
& + \sum_{i<j<k<l<m<p=1}^7 \hspace{-0.8cm}\eta_i\eta_j\eta_k\eta_l\eta_m\eta_p
\psi_-^{ijklmp(8)} - \eta_1\eta_2\eta_3\eta_4\eta_5\eta_6\eta_7
h_-^{1234567(8)}~.~~~\nonumber\\
\end{align}
~\\
For the $\mathcal{N}_G=6$ theory, we have the $(+2,+{3\over
2}^{6},+1^{15},+{1\over 2}^{20},0^{15},-{1\over
2}^{6},-1)$-supergravity multiplet diamond
\begin{align} 
\Phi^{\mathcal{N}_G=6} = \Phi^{\mathcal{N}_G=8}|_{\eta_7,\eta_8
\rightarrow 0}
=&{} h_+ + \sum_{i=1,2,3,4,5,6} \hspace{-0.3cm}\eta_i\psi_+^i +
\sum_{i<j=1,2,3,4,5,6}\hspace{-0.3cm} \eta_i\eta_j v_+^{ij}  +
\sum_{i<j<k=1,2,3,4,5,6}\hspace{-0.5cm} \eta_i\eta_j\eta_k \chi_+^{ijk}\nonumber\\
& +\sum_{i<j<k<l=1,2,3,4,5,6} \hspace{-0.5cm} \eta_i\eta_j\eta_k\eta_l
\phi^{ijkl}+\sum_{i<j<k<l<m=1,2,3,4,5,6}\hspace{-0.5cm}
\eta_i\eta_j\eta_k\eta_l\eta_m \chi_-^{ijklm}\nonumber\\
& + \eta_1\eta_2\eta_3\eta_4\eta_5\eta_6 v_-^{123456}~,~~~
\end{align}
and the $(+1,+{1\over 2}^{6},0^{15},-{1\over 2}^{20},-1^{15},-{3\over
2}^{6},-2)$ supergravity multiplet diamond
\begin{align} 
\Psi^{\mathcal{N}_G=6} = \int    d\eta_7 d\eta_8
\Phi^{\mathcal{N}_G=8} =&{}
-v_+^{(78)} -\sum_{i=1,2,3,4,5,6} \eta_i\chi_+^{i(78)}
-\sum_{i<j=1,2,3,4,5,6} \eta_i\eta_j \phi^{ij(78)}  \nonumber \\
&-\!\!\! \sum_{i<j<k=1,2,3,4,5,6} \eta_i\eta_j\eta_k \chi_-^{ijk(78)}
- \sum_{i<j<k<l=1,2,3,4,5,6} \hspace{-0.5cm} \eta_i\eta_j\eta_k\eta_l v_-^{ijkl(78)}
\nonumber \\
& -\!\!\! \sum_{i<j<k<l<m=1,2,3,4,5,6}\hspace{-0.8cm} \eta_i\eta_j\eta_k\eta_l\eta_m
\psi_-^{ijklm(78)} - \eta_1\eta_2\eta_3\eta_4\eta_5\eta_6
h_-^{123456(78)} ~.~~~\nonumber\\
\end{align}
We also have the $(+{3\over 2},+1^{6},+{1\over 2}^{15},0^{20},-{1\over
2}^{15},-1^{6},-{3\over 2})$ gravitino supermultiplet diamond
\begin{align}
\Theta^{\mathcal{N}_G=6} \equiv \int d
\eta_{4}\Phi^{\mathcal{N}_G=8} |_{\eta_8\rightarrow 0}
={}& \psi_+^{(4)} - \sum_{i=1,2,3,5,6,7} \eta_i v_+^{i(4)} +
\sum_{i<j=1,2,3,5,6,7} \eta_i\eta_j \chi_+^{ij(4)}  \nonumber \\
&-\!\!\!\sum_{i<j<k=1,2,3,5,6,7} \eta_i\eta_j\eta_k \phi^{ijk(4)} +
\sum_{i<j<k<l=1,2,3,5,6,7} \hspace{-0.5cm} \eta_i\eta_j\eta_k\eta_l \chi_-^{ijkl(4)}
\nonumber \\
&-\!\!\!\sum_{i<j<k<l<m=1,2,3,5,6,7} \hspace{-0.8cm}\eta_i\eta_j\eta_k\eta_l\eta_m
v_-^{ijklm(4)} + \eta_1\eta_2\eta_3\eta_5\eta_6\eta_7
\psi_-^{123567(4)}\,,
\end{align}
and its CPT-conjugate gravitino supermultiplet diamond
\begin{align}
\Gamma^{\mathcal{N}_G=6} \equiv \int
d\eta_{8}\Phi^{\mathcal{N}_G=8} |_{\eta_4\rightarrow 0}
={}& \psi_+^{(8)} - \sum_{i=1,2,3,5,6,7} \eta_i v_+^{i(8)}
+\sum_{i<j=1,2,3,5,6,7} \eta_i\eta_j \chi_+^{ij(8)}  \nonumber \\
&-\!\!\!\sum_{i<j<k=1,2,3,5,6,7} \eta_i\eta_j\eta_k \phi^{ijk(8)}
+\sum_{i<j<k<l=1,2,3,5,6,7} \hspace{-0.5cm} \eta_i\eta_j\eta_k\eta_l \chi_-^{ijkl(8)}
\nonumber \\
&-\!\!\!\sum_{i<j<k<l<m=1,2,3,5,6,7} \hspace{-0.8cm}\eta_i\eta_j\eta_k\eta_l\eta_m
v_-^{ijklm(8)} + \eta_1\eta_2\eta_3\eta_5\eta_6\eta_7
\psi_-^{123567(8)}\,. \end{align}
~\\
For the $\mathcal{N}_G=5$ theory, we have the $(+2,+{3\over
2}^5,+1^{10},+{1\over 2}^{10},0^{5},-{1\over
2})$ supergravity multiplet diamond
\begin{align}
\Phi^{\mathcal{N}_G=5} =
\Phi^{\mathcal{N}_G=8}|_{\eta_6,\eta_7,\eta_8 \rightarrow 0}
={}& h_+ + \sum_{i=1,2,3,4,5} \eta_i\psi_+^i +\sum_{i<j=1,2,3,4,5} \hspace{-0.3cm}\eta_i\eta_j v_+^{ij}
+\sum_{i<j<k=1,2,3,4,5}\hspace{-0.5cm} \eta_i\eta_j\eta_k \chi_+^{ijk}\nonumber\\
& +\sum_{i<j<k<l=1,2,3,4,5}  \eta_i\eta_j\eta_k\eta_l
\phi^{ijkl}+\eta_1\eta_2\eta_3\eta_4\eta_5 \chi_-^{12345}\,,
\end{align}
and the $(+{1\over 2},0^{5},-{1\over 2}^{10},-1^{10},-{3\over
2}^5,-2)$ supergravity multiplet diamond
\begin{align}
\Psi^{\mathcal{N}_G=5} = \int   \prod_{A=6}^8 d\eta_A
\Phi^{\mathcal{N}_G=8} ={}&
-\chi_+^{(678)} +\sum_{i=1,2,3,4,5} \eta_i\phi^{i(678)}  -\sum_{i<j=1,2,3,4,5} \eta_i\eta_j \chi_-^{ij(678)}  \nonumber \\
&+\sum_{i<j<k=1,2,3,4,5}\hspace{-0.5cm} \eta_i\eta_j\eta_k v_-^{ijk(678)} -\sum_{i<j<k<l=1,2,3,4,5} \hspace{-0.5cm} \eta_i\eta_j\eta_k\eta_l \psi_-^{ijkl(678)} \nonumber \\
& + \eta_1\eta_2\eta_3\eta_4\eta_5 h_-^{12345(678)} \,.
\end{align}
There is also a $(+{3\over 2},+1^{5},+{1\over 2}^{10},0^{10},-{1\over
2}^{5},-1)$ gravitino supermultiplet diamond
\begin{align}
\Theta^{\mathcal{N}_G=5}\equiv \int d
\eta_{4}\Phi^{\mathcal{N}_G=8} |_{\eta_7,\eta_8\rightarrow 0}
={}& \psi_+^{(4)} - \sum_{i=1,2,3,5,6} \eta_i v_+^{i(4)} +\sum_{i<j=1,2,3,5,6} \eta_i\eta_j\chi_+^{ij(4)} \nonumber \\
&-\sum_{i<j<k=1,2,3,5,6} \hspace{-0.5cm} \eta_i\eta_j\eta_k \phi_-^{ijk(4)} +\sum_{i<j<k<l=1,2,3,5,6}\hspace{-0.5cm} \eta_i\eta_j\eta_k\eta_l \chi_-^{ijkl(4)} \nonumber \\
&- \eta_1\eta_2\eta_3\eta_5\eta_6 \psi_-^{12356(4)}\,,
\end{align}
as well as the $(+1,+{1\over 2}^{5},0^{10},-{1\over 2}^{10},-1^{5},-{3\over
2})$ gravitino supermultiplet diamond
\begin{align}
\Gamma^{\mathcal{N}_G=5}\equiv \int
d\eta_{7}d\eta_{8}\Phi^{\mathcal{N}_G=8} |_{\eta_4\rightarrow 0}
={}& -v_+^{(78)} - \sum_{i=1,2,3,5,6} \eta_i \chi_+^{i(78)} -\sum_{i<j=1,2,3,5,6} \eta_i\eta_j\phi^{ij(78)} \nonumber \\
&-\sum_{i<j<k=1,2,3,5,6} \hspace{-0.5cm}\eta_i\eta_j\eta_k \chi_-^{ijk(78)} -\sum_{i<j<k<l=1,2,3,5,6} \hspace{-0.5cm}\eta_i\eta_j\eta_k\eta_l v_-^{ijkl(78)} \nonumber \\
& -\eta_1\eta_2\eta_3\eta_5\eta_6 \psi_-^{12356(78)} \,. \end{align}
~\\
For the $\mathcal{N}_G=4$ theory, we have the $(+2,+{3\over 2}^4,+1^6,+{1\over
2}^4,0)$ supergravity multiplet diamond
\begin{align}
\Phi^{\mathcal{N}_G=4} = \Phi^{\mathcal{N}_G=8}|_{\eta_5,
\eta_6,\eta_7,\eta_8 \rightarrow 0}
={}& h_+ + \sum_{i=1,2,3,4} \eta_i\psi_+^i +\sum_{i<j=1,2,3,4} \eta_i\eta_j v_+^{ij}  \nonumber \\
&+\sum_{i<j<k=1,2,3,4} \eta_i\eta_j\eta_k \chi_+^{ijk} +
\eta_1\eta_2\eta_3\eta_4 \phi^{1234} \,, \end{align}
and the $(0,-{1\over 2}^4,-1^6,-{3\over
2}^4,-2)$ supergravity multiplet diamond
\begin{align}
\Psi^{\mathcal{N}_G=4} = \int   \prod_{A=5}^8 d\eta_A
\Phi^{\mathcal{N}_G=8} ={}&
\phi^{(5678)} + \sum_{i=1,2,3,4} \eta_i\chi_-^{i(5678)}  +\sum_{i<j=1,2,3,4} \eta_i\eta_j v_-^{ij(5678)}  \nonumber \\
&+\sum_{i<j<k=1,2,3,4} \eta_i\eta_j\eta_k \psi_-^{ijk(5678)} +
\eta_1\eta_2\eta_3\eta_4 h_-^{1234(5678)} \,.
\end{align}
There is also a $(+{3\over 2},+1^4,+{1\over 2}^6,0^4,-{1\over
2})$ gravitino supermultiplet diamond
\begin{align}
\Theta^{\mathcal{N}_G=4} \equiv \int  d
\eta_{4}\Phi^{\mathcal{N}_G=8} |_{\eta_6,\eta_7,\eta_8\rightarrow 0}
={}& \psi_+^{(4)} - \sum_{i=1,2,3,5}\eta_iv_+^{i(4)} + \sum_{i<j=1,2,3,5} \eta_i\eta_j\chi_+^{ij(4)} \nonumber \\
& -\sum_{i<j<k=1,2,3,5} \eta_i\eta_j\eta_k \phi^{ijk(4)} +
\eta_1\eta_2\eta_3\eta_5 \chi_-^{1235(4)} \,,
\end{align}
and a $(+{1\over 2},0^4,-{1\over 2}^6,-1^4,-{3\over
2})$ gravitino supermultiplet diamond
\begin{align}
\Gamma^{\mathcal{N}_G=4} \equiv \int
d\eta_{6}d\eta_{7}d\eta_{8}\Phi^{\mathcal{N}_G=8}
|_{\eta_4\rightarrow 0}
={}&  -\chi_+^{(678)} + \sum_{i=1,2,3,5} \eta_i \phi^{i(678)} -\sum_{i<j=1,2,3,5} \eta_i\eta_j\chi_-^{ij(678)} \nonumber \\
&+\sum_{i<j<k=1,2,3,5} \eta_i\eta_j\eta_k v_-^{ijk(678)} -
\eta_1\eta_2\eta_3\eta_5 \psi_-^{1235(678)} \,, \end{align}
as well as a $(+1,+{1\over 2}^4,0^6,-{1\over 2}^4,-1)$ vector multiplet
diamond
\begin{align}
\Theta^{\mathcal{N}_G=4}_{vector} \equiv \int  d\eta_3 d
\eta_{4}\Phi^{\mathcal{N}_G=8} |_{\eta_7,\eta_8\rightarrow 0}
={}& -v_+^{(34)} - \sum_{i=1,2,5,6} \eta_i \chi_+^{i(34)} -\sum_{i<j=1,2,5,6} \eta_i\eta_j\phi^{ij(34)} \nonumber \\
&-\sum_{i<j<k=1,2,5,6} \eta_i\eta_j\eta_k \chi_-^{ijk(34)} -
\eta_1\eta_2\eta_5\eta_6 v_-^{1256(34)}\,,
\end{align}
with its CPT-conjugate
\begin{align}
\Gamma^{\mathcal{N}_G=4}_{vector} \equiv \int
d\eta_{7}d\eta_{8}\Phi^{\mathcal{N}_G=8} |_{\eta_3,\eta_4\rightarrow
0}
={}& -v_+^{(78)} - \sum_{i=1,2,5,6} \eta_i \chi_+^{i(78)} -\sum_{i<j=1,2,5,6} \eta_i\eta_j\phi^{ij(78)} \nonumber \\
&-\sum_{i<j<k=1,2,5,6} \eta_i\eta_j\eta_k \chi_-^{ijk(78)} -
\eta_1\eta_2\eta_5\eta_6 v_-^{1256(78)} \,. \end{align}
~\\
For the $\mathcal{N}_G=3$ theory, we have the $(+2,+{3\over 2}^3,+1^3,+{1\over
2})$ supergravity multiplet diamond
\begin{align}
\Phi^{\mathcal{N}_G=3} = \Phi^{\mathcal{N}_G=8}|_{\eta_4, \eta_5,
\eta_6,\eta_7,\eta_8 \rightarrow 0} ={}& h_+ + \sum_{i=1,2,3}
\eta_i\psi_+^i + \sum_{i<j=1,2,3} \eta_i\eta_j v_+^{ij} +
\eta_1\eta_2\eta_3 \chi_+^{123} \,, \end{align}
and the $(-{1\over 2},-1^3,-{3\over 2}^3,-2)$ supergravity multiplet
diamond
\begin{align} \Psi^{\mathcal{N}_G=3} =& \int   \prod_{A=4}^8 d\eta_A
\Phi^{\mathcal{N}_G=8} \nonumber\\={} &\chi_-^{(45678)} -
\sum_{i=1,2,3} \eta_i v_-^{i(45678)} + \sum_{i<j=1,2,3} \eta_i\eta_j
\psi_-^{ij(45678)} - \eta_1\eta_2\eta_3 h_-^{123(45678)}\,.
\end{align}
Again there is also a $(+{3\over 2},+1^3,+{1\over
2}^3,0)$ gravitino supermultiplet diamond
\begin{align}
\Theta^{\mathcal{N}_G=3} \equiv &\int  d
\eta_{4}\Phi^{\mathcal{N}_G=8}
|_{\eta_5,\eta_6,\eta_7,\eta_8\rightarrow 0} \nonumber\\
={}& \psi_+^ {(4)} - \sum_{i=1,2,3} \eta_i v_+^{i(4)} +
\sum_{i<j=1,2,3} \eta_i\eta_j \chi_+^{ij(4)} - \eta_1\eta_2\eta_3
\phi^{123(4)} \,,
\end{align}
and a $(0,-{1\over 2}^3,-1^3,-{3\over 2})$ gravitino supermultiplet
diamond
\begin{align}
\Gamma^{\mathcal{N}_G=3} \equiv &\int d\eta_5
d\eta_{6}d\eta_{7}d\eta_{8}\Phi^{\mathcal{N}_G=8}
|_{\eta_4\rightarrow 0} \nonumber\\
={}&  \phi^{(5678)} + \sum_{i=1,2,3} \eta_i \chi_-^{i(5678)} +
\sum_{i<j=1,2,3} \eta_i\eta_j v_-^{ij(5678)} + \eta_1\eta_2\eta_3
\psi_-^{123(5678)} \,.
\end{align}
And there is a $(+1,+{1\over 2}^3,0^3,-{1\over
2})$ vector multiplet diamond
\begin{align}
\Theta^{\mathcal{N}_G=3}_{vector} \equiv &\int  d\eta_3 d
\eta_{4}\Phi^{\mathcal{N}_G=8} |_{\eta_6,\eta_7,\eta_8\rightarrow
0}\nonumber\\
 ={}& -v_+^{(34)} - \sum_{i=1,2,5} \eta_i
\chi_+^{i(34)} - \sum_{i<j=1,2,5} \eta_i\eta_j\phi^{ij(34)} -
\eta_1\eta_2\eta_5 \chi_-^{125(34)} \,,
\end{align}
and a $(+{1\over 2},0^3,-{1\over 2}^3,-1)$ vector multiplet
diamond
\begin{align}
\Gamma^{\mathcal{N}_G=3}_{vector} \equiv &\int
d\eta_{6}d\eta_{7}d\eta_{8}\Phi^{\mathcal{N}_G=8}
|_{\eta_3,\eta_4\rightarrow 0} \nonumber\\
={}& -\chi_+^{(678)} + \sum_{i=1,2,5} \eta_i \phi^{i(678)}  -
\sum_{i<j=1,2,5}\eta_i\eta_j \chi_-^{ij(678)}+ \eta_1\eta_2\eta_5
v_-^{125(678)} \,. \end{align}
~\\
For the $\mathcal{N}_G=2$ theory, we have the $(+2,+{3\over
2}^2,+1)$ supergravity multiplet diamond
\begin{align}
\Phi^{\mathcal{N}_G=2}&=
\Phi^{\mathcal{N}_G=8}|_{\eta_{2},\eta_3,\eta_4,
\eta_6,\eta_7,\eta_8 \rightarrow 0} = h_+ + \eta_1\psi_+^1 +
\eta_5\psi_+^5 + \eta_1\eta_5 v_+^{15} \,, \end{align}
and the $(-1,-{3\over 2}^2,-2)$ supergravity multiplet diamond
\begin{align}
\Psi^{\mathcal{N}_G=2} =& \int \prod_{A=2}^4 d\eta_A \prod_{A=6}^8
d\eta_A \Phi^{\mathcal{N}_G=8} \nonumber\\
= &-v_-^{(234678)} - \eta_1\psi_-^{1(234678)} -
\eta_5\psi_-^{5(234678)} - \eta_1\eta_5 h_-^{15(234678)}\,.
\end{align}
There is also a $(+1,+{1\over 2}^2,0)$ vector multiplet diamond
\begin{align}
\Theta^{\mathcal{N}_G=2}_{vector} &\equiv \int  d\eta_3 d\eta_4
\Phi^{\mathcal{N}_G=8} |_{\eta_5,\ldots,\eta_8\rightarrow 0} =
-v_+^{(34)} - \eta_1\chi_+^{1(34)} - \eta_2\chi_+^{2(34)} -
\eta_1\eta_2 \phi^{12(34)} \,,
\end{align}
and a $(0,-{1\over 2}^2,-1)$ vector multiplet diamond
\begin{align}
\Gamma^{\mathcal{N}_G=2}_{vector} &\equiv \int \prod_{a=5}^8 d
\eta_{a}\Phi^{\mathcal{N}_G=8} |_{\eta_3,\eta_4\rightarrow 0} =
\phi^{(5678)} + \eta_1\chi_-^{1(5678)} + \eta_2\chi_-^{2(5678)} +
\eta_1\eta_2 v_-^{12(5678)} \,. \end{align}
Besides this vector supermultiplet, there is also a $(+{1\over 2},0^2,-{1\over
2})$ hypermultiplet diamond
\begin{align}
\Theta^{\mathcal{N}_G=2}_{hyper} &\equiv \int \prod_{a=2}^4 d
\eta_{a}\Phi^{\mathcal{N}_G=8} |_{\eta_6,\ldots,\eta_8\rightarrow 0}
= -\chi_+^{(234)} + \eta_1\phi^{1(234)} + \eta_5 \phi^{5(234)} -
\eta_1\eta_5 \chi_-^{15(234)}\,,
\end{align}
and its CPT-conjugate partner
\begin{align}
\Gamma^{\mathcal{N}_G=2}_{hyper} &\equiv \int \prod_{a=6}^8 d
\eta_{a}\Phi^{\mathcal{N}_G=8} |_{\eta_2,\ldots,\eta_4\rightarrow 0}
= -\chi_+^{(678)}+ \eta_1\phi^{1(678)} + \eta_5\phi^{5(678)} -
\eta_1\eta_5 \chi_-^{15(678)}\,. \end{align}
~\\
For the $\mathcal{N}_G=1$ theory, we have the $(+2,+{3\over 2})$ and
$(-{3\over 2},-2)$ supergravity multiplet diamond
\begin{align}
\Phi^{\mathcal{N}_G=1} &= \Phi^{\mathcal{N}_G=8}|_{\eta_{2},\ldots,\eta_8 \rightarrow 0} = h_+ + \eta_1\psi_+^1 \,, \\
\Psi^{\mathcal{N}_G=1} &= \int \prod_{A=2}^8 d\eta_A
\Phi^{\mathcal{N}_G=8} = -\psi_-^{(2345678)} + \eta_1
h_-^{1(2345678)}\,,
\end{align}
as well as the $(+{1\over 2},0)$ and $(0,-{1\over
2})$ chiral supermultiplet diamond
\begin{align}
\Theta_{chiral}^{\mathcal{N}_G=1} &\equiv \int \prod_{a=2}^4 d
\eta_{a}\Phi^{\mathcal{N}_G=8} |_{\eta_5,\ldots,\eta_8\rightarrow 0}
= -\chi_+^{(234)} + \eta_1\phi^{1(234)}~,~~~\\
\Gamma_{chiral}^{\mathcal{N}_G=1} &\equiv \int \prod_{a=5}^8 d
\eta_{a}\Phi^{\mathcal{N}_G=8} |_{\eta_2,\ldots,\eta_4\rightarrow 0}
= \phi^{(5678)} + \eta_1 \chi_-^{1(5678)}\,.
\end{align}
~\\
Finally, for the theory without any supersymmetry (the ``$\mathcal{N}_G=0$'' theory),
the ``superfield'' contains only one single state. This is the graviton
\begin{align}
\Phi^{\mathcal{N}_G=0} &= \Phi^{\mathcal{N}_G=8}|_{\eta_{1},\ldots,\eta_8 \rightarrow 0} = h_+\,, \\
\Psi^{\mathcal{N}_G=0} &= \int \prod_{A=1}^8 d\eta_A
\Phi^{\mathcal{N}_G=8} = h_-^{(12345678)}\,, \end{align}
as well as the two scalars discussed at length in the text,
\begin{align}
\Theta^{\mathcal{N}_G=0} &= \int \prod_{a=1}^4
d\eta_{a}\Phi^{\mathcal{N}_G=8} |_{\eta_5,\ldots,\eta_8\rightarrow
0} = \phi^{(1234)}\,,\\
\Gamma^{\mathcal{N}_G=0} &= \int \prod_{a=5}^8
d\eta_{a}\Phi^{\mathcal{N}_G=8} |_{\eta_1,\ldots,\eta_4\rightarrow
0} = \phi^{(5678)}\,.
\end{align}

\end{document}